\documentclass[preprint,preprintnumbers,amsmath,amssymb,floatfix,superscriptaddress]{revtex4-1}

\usepackage{graphicx}
\usepackage{dcolumn}
\usepackage{bm}
\usepackage{longtable}

\usepackage{color}
\definecolor{lightblue}{rgb}{0.2,0.2,0.7}
\definecolor{darkblue}{rgb}{0,0.25,0.5}
\definecolor{redbrown}{rgb}{0.875,0.25,0.125}
\definecolor{darkgreen}{rgb}{0,0.5,0}

\usepackage{hyperref}

\newcommand{\bra}[1]{\ensuremath{\langle #1 \vert}}
\newcommand{\ket}[1]{\ensuremath{\vert #1  \rangle}}
\newcommand{\braket}[2]{\ensuremath{\langle  #1 \vert #2  \rangle}}

\renewcommand{\b}[1]{\ensuremath{\mathbf{#1}}}

\newcommand{\T}{\ensuremath{\text{T}}}
\renewcommand{\L}{\ensuremath{\text{L}}}

\newcommand{\lin}{\ensuremath{\text{lin}}}

\newcommand{\Psib}{\ensuremath{\overline{\Psi}}}

\newcommand{\FN}{\ensuremath{\text{FN}}}

\renewcommand{\L}{\ensuremath{\text{L}}}
\renewcommand{\d}{\ensuremath{\text{d}}}
\newcommand{\p}{\ensuremath{\text{p}}}
\renewcommand{\L}{\ensuremath{\text{L}}}
\newcommand{\R}{\ensuremath{\text{R}}}

\begin{document}

\title{The Valence-Bond Quantum Monte Carlo Method}

\author{Slavko Radenković} 
\email{slavkoradenkovic@kg.ac.rs} 
\affiliation{Faculty of Science, University of Kragujevac, 34000 Kragujevac, Serbia} 

\author{Dominik Domin} 
\email{dominik.domin@universite-paris-saclay.fr} 
\affiliation{Université Paris-Saclay, CNRS, Institut de Chimie Physique, UMR8000, 91405 Orsay, France} 

\author{Julien Toulouse} 
\email{toulouse@lct.jussieu.fr} 
\affiliation{Laboratoire de Chimie Th\'eorique, Sorbonne Universit\'e and CNRS, F-75005 Paris, France} 
\affiliation{Institut Universitaire de France, F-75005 Paris, France}

\author{Benoît Braïda } 
\email{braida@lct.jussieu.fr} 
\affiliation{Laboratoire de Chimie Th\'eorique, Sorbonne Universit\'e and CNRS, F-75005 Paris, France}

\date{July 28, 2022}

\maketitle
\tableofcontents

\section*{Abstract}
The VB-QMC method is presented in this chapter. It consists of using in quantum Monte Carlo (QMC) approaches with a wave function expressed as a usually short expansion of classical Valence-Bond (VB) structures supplemented by a Jastrow factor to account for dynamical correlation. Two variants exist: the VB-VMC (using variational Monte Carlo) and VB-DMC (using diffusion Monte Carlo) methods. QMC algorithms circumvent the notorious non-orthogonality issue of classical VB approaches, and allow highly efficient calculations on massively parallel machines. Calculation of VB weights and resonance energies are possible at the VB-VMC level, which makes VB-VMC a correlated method retaining all the interpretative capabilities of classical VB methods.  Several recent applications are shown to illustrate the potential of this method as a modern alternative to classical VB methods to study ground and excited states of molecules.

\section*{Keywords}
Valence Bond; non-orthogonal wave functions; chemical bonding; quantum Monte Carlo; variational Monte Carlo; diffusion Monte Carlo; interpretative methods; structure weights; resonance energies; excited states; hypervalency; charge-shift bonding; three-electron bond.

\section*{Glossary}
\begin{itemize}
\item \textbf{BOVB} Breathing-orbital Valence Bond, a VB method in which each VB structure is allowed to have a different set of orbitals.
\item \textbf{CASPT2} Complete-active-space second-order perturbation theory, a standard wave-function method for both static and dynamical correlation.
\item \textbf{CASSCF} Complete-active-space self-consistent field, a standard wave-function method for static correlation.
\item \textbf{CI} Configuration interaction, a standard wave-function method.
\item \textbf{DMC} Diffusion Monte Carlo, a variant of QMC extracting the best variational wave function having the same nodes as the trial wave function.
\item \textbf{Jastrow factor} A function explicitly depending on the interparticle coordinates to introduce dynamical electron correlation in the wave function.
\item \textbf{JVB} Jastrow-Valence-Bond wave function, a wave function consisting in a Jastrow factor multiplied by an expansion in VB structures.
\item \textbf{Jastrow-Slater wave function} A form of wave function consisting in a Jastrow factor multiplied by an expansion in Slater determinants.
\item \textbf{p-BOVB} Partial BOVB, a variant of BOVB in which different VB structures share the same set of orbitals.
\item \textbf{QMC} Quantum Monte Carlo, a family of stochastic methods for computational quantum mechanics.
\item \textbf{SD-BOVB} Split-delocalized BOVB, BOVB variant in which the doubly occupied active orbitals are split into two spin-coupled orbitals and the inactive orbitals are delocalized.
\item \textbf{VB-DMC} Valence Bond Diffusion Monte Carlo, variant of VB-QMC using the DMC method.
\item \textbf{VB-QMC} Valence Bond Quantum Monte Carlo, the method presented in this chapter which uses JVB wave functions in QMC algorithms.
\item \textbf{VBSCF} Valence Bond self-consistent field, a VB method in which both the coefficients of the VB structures and the common set of orbitals are optimized.
\item \textbf{VB-VMC} Valence Bond Variational Monte Carlo, variant of VB-QMC using the VMC method.
\item \textbf{VB structures} A linear combination of Slater determinants made of non-orthogonal orbitals and representing a Lewis chemical structure.
\item \textbf{VMC} Variational Monte Carlo, the simplest variant of QMC where multidimensional integrals are calculated by sampling the probability density of a trial wave function. 
 
\end{itemize}

\section*{Key Points}
\begin{itemize}
    \item Description of the VB-QMC method: Jastrow-VB wave functions, VMC and DMC algorithms, optimization of the wave-function parameters. 
    \item Definition of the weights of the VB structures in the VB-QMC method.
    \item Benchmark of the VB-QMC method on a selection of small molecules.
    \item Applications of the VB-QMC method: V state of ethylene, (DTCNE)$_2$ ``pancake bonding'' prototype, XeF$_2$ prototype.
\end{itemize}
.

\section{Introduction}

Monte Carlo (MC) integration is a family of stochastic algorithms in which repeated random sampling is used to numerically approximate multi-dimensional integrals.~\cite{Allen-Tildesley}. Quantum Monte Carlo (QMC) methods~\cite{Luchow-QMC-WIREs-2011, Austin-QMC-ChemRev-2012,TouAssUmr-INC-16} correspond to the application of MC numerical techniques in the context of quantum mechanics to solve the many-electron Schr\"odinger equation, and they represent an alternative to standard quantum-chemistry methods. Since the integrals are estimated with numerical MC sampling rather than by analytical integration, far more flexibility is possible in the form of the wave function. In particular, sophisticated wave functions featuring an explicit electronic correlation function (often referred to as a ``Jastrow factor'') are typically used and this allows for a very large percentage of the electronic correlation energy to be retrieved. Furthermore, the so-called projector QMC methods are capable of calculating very accurate ground-state energies even with trial wave functions with relatively modest complexity.

The computational cost of most QMC algorithms scales favorably with the number of electrons. In general they scale cubically with the number of electrons~\cite{AssMorFil-JCTC-17} and this can be improved up to linear scaling for very large systems. On the downside, the statistical error of computed quantities in MC simulations
decreases slowly, in $1/\sqrt{M}$, where $M$ is the number of sampling points. This is the primary reason why QMC simulations tend to be computationally expensive when compared to deterministic methods such as density-functional theory (DFT). So while QMC calculations have a similar scaling with system size as DFT the prefactor in the scaling of QMC calculations is orders of magnitude larger. However, QMC methods are often more accurate than DFT with usual approximate density functionals. They tend to be on par or more accurate than very large coupled-cluster calculations (CCSD(T), CCSDTQ, ...) which scale far less favorably with system size both in terms of memory and in terms of computational effort. Furthermore, QMC methods are massively parallelizable since the MC sampling is inherently distributable on hundreds of thousands of processor cores with near 100\% efficiency.

QMC techniques have the same computational cost for evaluating wave functions built with non-orthogonal orbitals as for those built with orthogonal orbitals. This is a marked difference with standard quantum-chemistry methods based on analytical integration, where non-orthogonality immensely complexifies the resolution of the Schr\"odinger equation, and has slowed down for decades the development of \textit{ab initio} valence bond (VB) methods.  Another reason to use QMC techniques in conjunction with VB theory is that while the VB ansatz allows for a direct  description of the electron structure in terms of Lewis structures (or VB structures), traditional VB methods cannot fully account for the dynamical electronic correlation. In particular, the breathing-orbital valence bond (BOVB) method~\cite{HibHumByrLen-JCP-94,HibSha-TCA-02}, which is the most used correlated classical VB method in chemical applications (see corresponding chapter in this book) only accounts for partial inclusion of the dynamical correlation, and is inherently limited to an expansion of only a handful of VB structures. By contrast, using QMC algorithms allows the calculations of non-orthogonal VB wave functions that include a larger amount of dynamical correlation through an explicit Jastrow correlation function and that can include hundreds (or even thousands in the case of small molecules) VB structures.

The most commonly employed QMC methods are variational Monte Carlo (VMC) and diffusion Monte Carlo (DMC). The VMC method uses a flexible trial wave function which consists of a Jastrow correlation factor multiplied by a single or a linear combination of Slater determinants and applies MC numerical techniques for optimizing the wave-function parameters and calculating energies and properties. The DMC method goes beyond VMC by projecting the previously optimized VMC trial wave function onto the exact ground state through a stochastic implementation of the power method. In practice, the DMC method is applied within the fixed-node (FN) approximation, which results in extracting the best variational wave function having the same nodes as the trial wave function. These QMC methods have been successfully applied to study the electronic structure of molecules and solids where electronic correlation plays an important role,~\cite{WagnerMitas-TM-JCP-2007, Wagner-TM-JCTC-2016, Matxain-Sc-dimer-JCP-2008, Purwanto-Cr-JCP-2015} and there is a continuous extension of the applications of the VMC and DMC methods thanks to improved wave-function optimization techniques~\cite{TouUmr-JCP-07,UmrTouFilSorHen-PRL-07,TouUmr-JCP-08} and to growing computational resources. Generally, the VMC and DMC methods require modest amounts of computer memory and can be efficiently parallelized, which make them ideally suited for massively parallel supercomputers.

It has been shown that the accuracy of the VMC and DMC methods strongly depends on the reliability of the employed trial wave function.~\cite{Clay-JCP-2015} The most straightforward approach to obtain a reliable trial wave function is to use a Jastrow-Slater multi-determinant wave function expanded into delocalized molecular orbitals. It has been shown indeed that a Jastrow factor multiplied by a complete-active-space (CAS)~\cite{TouUmr-JCP-08,Zimmerman-ExcMethylene-JCP-2009} or a more general truncated configuration-interaction (CI)~\cite{Morales-JCTC-2012,Giner-MolPhys-2016} expansion of Slater determinants is capable of providing chemical accuracy. However, a major challenge when using such Jastrow-Slater multi-determinant wave functions is how to systematically select the best Slater determinants entering the trial wave function. 

In this context, the family of VB methods~\cite{ShaHib-BOOK-08} constitutes an interesting alternative
to methods based on delocalized orbitals. “Classical” VB wave functions use non-orthogonal orbitals strictly localized on atoms, which leads to determinantal expansions that are more compact than those obtained from delocalized orbitals. Moreover, this brings strong interpretative capabilities since these VB wave functions represent a superposition of specific Lewis chemical structures~\cite{Wu-VB-ChemRev-2011, shurki-VBChaptert-2021}. Following a preliminary study in which BOVB wave functions were tested in DMC~\cite{Domin-BOVB-QMC-JPCA-2008}, general Jastrow-Valence-Bond (JVB) wave functions made of a Jastrow factor multiplied by compact VB wave functions based on non-orthogonal orbitals strictly localized on atoms were introduced in QMC in Ref.~\onlinecite{Braida-VBQMC-JCP-2011}. In contrast to standard VB techniques using deterministic algorithms, the use of non-orthogonal orbitals do not cause any additional computational costs and algorithm complexity in QMC. We refer to this approach as VB-QMC in the remainder of this chapter. More specifically, two variants exist: the VB-VMC and VB-DMC methods. In the VB-VMC method, these JVB wave functions are used in VMC and the wave function parameters (Jastrow parameters, coefficients of the determinants, and coefficients of the orbitals) are optimized using the linear energy minimization method~\cite{TouUmr-JCP-07,UmrTouFilSorHen-PRL-07,TouUmr-JCP-08}. In the VB-DMC method, the previously optimized JVB wave functions are used in DMC. This VB-QMC approach has produced promising results for strongly correlated systems.~\cite{Braida-VBQMC-JCP-2011} This chapter provides a review of this VB-QMC approach, with a detailed technical presentation of the method, followed by a short review of already published applications using it. 

We shall mention that there are several other approaches using non-orthogonal multi-determinant expansions that have been used in QMC. Anderson and Goddard III~\cite{Anderson-GVB-QMC-JCP-2010} used perfect-pairing generalized Valence Bond (GVB) wave functions in QMC. Fracchia \textit{et al.}~\cite{Fracchia-GVB-QMC-JCTC-2012} developed a new type of Jastrow-Slater wave functions based on a simplified form of the GVB wave function. The idea behind the latter development was to be able to deal with large systems. Pathak and Wagner~\cite{Pathak-multideterm-QMC-JCP-2018} considered BOVB-like wave functions where each determinant has a different set of orbitals. Landinez Borda \textit{et al.}~\cite{Landinez_multideterm-QMC-JCP-2019} used non-orthogonal multi-determinant wave functions in auxiliary-field QMC. Note that these latter two developments were for quantitative calculations, not for interpretative purposes.

\section{Valence-Bond Quantum Monte Carlo methodology}

\subsection{Jastrow-Valence-Bond wave functions}

In VB-QMC, we use Jastrow-Valence-Bond (JVB) wave functions of the form~\cite{BraTouCafUmr-JCP-11}
\begin{eqnarray}
\ket{\Psi_0} = \hat{J} \, \sum_{I=1}^{N_\text{VB}} c_I \ket{\Phi_I},
\label{PsiJVBSCF}
\end{eqnarray}
where $\hat{J}$ is a Jastrow correlation-factor operator and $\{\ket{\Phi_I}\}$ are $N_\text{VB}$ VB structures with real-valued coefficients $\{c_I\}$. Each VB structure is determined by a choice of an orbital configuration (or orbital occupation) and a spin coupling of these orbitals. We consider VB structures of the form (disregarding normalization)
\begin{eqnarray}
\ket{\Phi_I} &=& \prod_p^{\text{inactive}} \hat{a}^\dag_{p\uparrow} \hat{a}^\dag_{p\downarrow}
\prod_{(ij)}^{\substack{\text{active}\\\text{pairs}}} \left( \hat{a}^\dag_{i\uparrow} \hat{a}^\dag_{j\downarrow} -
\hat{a}^\dag_{i\downarrow} \hat{a}^\dag_{j\uparrow} \right) 
\prod_q^{\substack{\text{active}\\\text{unpaired}}} \hat{a}^\dag_{q\uparrow} \ket{\text{vac}},
\label{PhiI}
\end{eqnarray}
where $\hat{a}^\dag_{p\sigma}$ ($\sigma \in \{\uparrow,\downarrow\}$) is a spin-orbital creation operator and $\ket{\text{vac}}$ is the vacuum state of second quantization. The VB structures are thus made of inactive (always closed-shell) orbitals $p$,
spin-singlet pairs of active orbitals $(ij)$, and possibly remaining unpaired spin-up active orbitals $q$. We use inactive
orbitals that are either localized (expanded on the basis functions centered on a single atom), e.g. for core orbitals, or
delocalized (expanded on all the basis functions of all the atoms), e.g. for bonds made of inactive orbitals that do not mix
with the active orbitals. We use active orbitals that are always localized on a single atom, and they are typically identified
with valence atomic (hybrid) orbitals. Note that Eq.~(\ref{PhiI}) encompasses the case of
spin-singlet pairing of an active orbital with itself, i.e. $i=j$ giving simply $\hat{a}^\dag_{i\uparrow}
\hat{a}^\dag_{i\downarrow} - \hat{a}^\dag_{i\downarrow} \hat{a}^\dag_{i\uparrow}=2\hat{a}^\dag_{i\uparrow}
\hat{a}^\dag_{i\downarrow}$. The spin-coupling scheme based on singlet pairing used in Eq.~(\ref{PhiI}) is usually referred to as
the Heitler-London-Slater-Pauling (HLSP) scheme. There exist other spin-coupling schemes, but the HLSP scheme has the advantage
of providing a clear correspondence between each VB structure and a Lewis chemical structure, two singlet-paired active orbitals
representing either a bond or a lone pair. In principle, considering all possible pairings exhausts, for a given orbital
configuration, all the spin eigenstates (or spin couplings) of fixed quantum numbers
$S=N_{\text{unpaired}}/2$ and $M_S=+S$ ($N_{\text{unpaired}}$ is the number of spin-up unpaired electrons). In fact, considering
all possible pairings leads to an overcomplete set of spin couplings, but they can be reduced to a complete basis of
(non-redundant) spin couplings, also called Rumer basis~\cite{ShaHib-BOOK-08,Rumer-32,Pauncz-BOOK-79}. 
For most practical applications, only a small number of chemically relevant VB structures are kept in the calculation.

In practice, the VB structure $\ket{\Phi_I}$ can be expanded in $N_{\text{det},I}=2^{N_{\text{pairs}}}$ Slater determinants, 
\begin{eqnarray}
\ket{\Phi_I}=\sum_{k=1}^{N_{\text{det},I}} d_{I,k} \ket{D_k},
\end{eqnarray}
where $N_{\text{pairs}}$ is the number of pairs of different active orbitals ($i\not=j$) in this VB structure. The coefficients of the determinants $d_{I,\mu}$ for a given VB structure are all equal in absolute value. These Slater determinants are made of non-orthogonal spatial orbitals which are expanded on a maximum of $M$ basis functions $\{ \ket{\chi_\mu} \}$
\begin{eqnarray}
\ket{\phi_p} = \sum_{\mu=1}^M \lambda_{p,\mu} \ket{\chi_\mu},
\end{eqnarray}
where $\lambda_{p,\mu}$ are the orbital coefficients. For localized orbitals, many of such coefficients are in fact zero. In our different applications of the VB-QMC method, we have used for all-electron calculations Slater-type-orbital basis sets (which are adequate for reproducing the electron-nuclei cusp conditions), while for calculations using pseudopotentials we have used Gaussian-type-orbital basis sets.

The Jastrow factor operator $\hat{J}$ is a local operator, whose expression in position-spin representation consists of the exponential of sum of electron-nucleus, electron-electron, and electron-electron-nucleus terms~\cite{Umr-UNP-XX,GucSanUmrJai-PRB-05}
\begin{eqnarray}
J(\b{X}) = \exp( f_\text{en}(\b{X}) + f_\text{ee}(\b{X}) + f_\text{een}(\b{X})),
\label{}
\end{eqnarray}
where $\b{X}=(\b{x}_1,\b{x}_2,...,\b{x_N})$ designates the position-spin coordinates of $N$ electrons with $\b{x}_i=(\b{r}_i,\sigma_i)$ where $\b{r}_i \in \mathbb{R}^3$ and $\sigma_i \in \{\uparrow,\downarrow\}$. The terms $f_\text{en}(\b{X})$, $f_\text{ee}(\b{X})$, and $f_\text{een}(\b{X})$ are written as systematic expansions
\begin{eqnarray}
f_\text{en}(\b{X}) = \sum_{i=1}^N \sum_{\alpha=1}^{N_\text{nucl}} \left( \frac{a_{1,\alpha} R(r_{i\alpha})}{1+a_{2,\alpha} R(r_{i\alpha})} + \sum_{p=2}^{5} a_{p+1,\alpha} \; R(r_{i\alpha})^p \right),
\label{}
\end{eqnarray}
\begin{eqnarray}
f_\text{ee}(\b{X}) = \sum_{i=1}^N \sum_{j=i+1}^{N} \left(  \frac{b_1^{\sigma_i,\sigma_j} R(r_{ij})}{1+b_2 R(r_{ij})} + \sum_{p=2}^{5} b_{p+1} \; R(r_{ij})^p \right),
\label{}
\end{eqnarray}
\begin{eqnarray}
f_\text{een}(\b{X}) = \sum_{i=1}^N \sum_{j=i+1}^{N} \sum_{\alpha=1}^{N_\text{nucl}} \sum_{p=2}^{5} \sum_{k=0}^{p-1} \sum_{l=0}^{l_\text{max}} c_{p,k,l,\alpha} \bar{R}(r_{ij})^k [ \bar{R}(r_{i\alpha})^l  + \bar{R}(r_{j\alpha})^l] [\bar{R}(r_{i\alpha})\bar{R}(r_{j\alpha})]^m,
\label{}
\end{eqnarray}
where $l_\text{max}$ is $p-k$ if $k\not=0$ and $p-k-2$ if $k=0$, and only terms for which $m=(p-k-l)/2$ is an integer are included. In these expressions, $N_\text{nucl}$ is the number of nuclei, $r_{i\alpha}$ is the distance between electron $i$ and nucleus $\alpha$, $r_{ij}$ is the distance between electrons $i$ and $j$, $R(r) = r/(1+\kappa r)$ and $\bar{R}(r) = 1/(1+\kappa r)$ are scaling functions (depending on the fixed parameter $\kappa=0.8$). The coefficients $a_{p,\alpha}$ and $c_{p,k,l,\alpha}$ are constrained to be the same for nuclei $\alpha$ of the same chemical element. The coefficient $a_{1,\alpha}$ may be used to impose the electron-nuclei cusp conditions, but in the present work it is always zero since the electron-nuclei cusp conditions are already included in the orbitals for all-electron calculations or not present at all for calculations with pseudopotentials. The coefficients $b_1^{\uparrow,\downarrow}=1/2$ and $b_1^{\uparrow,\uparrow}=b_1^{\downarrow,\downarrow}=1/4$ are fixed to impose the electron-electron cusp condition. Some coefficients $c_{p,k,l,\alpha}$ are fixed in order not to alter the electron-nuclei and electron-electron cusp conditions.

\subsection{Optimization of the wave-function parameters}

For optimizing the parameters in the JVB wave functions, we adopt the following parametrization~\cite{TouUmr-JCP-08,BraTouCafUmr-JCP-11}
\begin{eqnarray}
\ket{\Psi(\b{p})} = \hat{J}(\bm{\alpha}) \, e^{\hat{\kappa}(\bm{\kappa})}\sum_{I=1}^{N_\text{VB}} c_I \ket{\Phi_I},
\label{PsiJVBp}
\end{eqnarray}
where $e^{\hat{\kappa}(\bm{\kappa})}$ is an orbital rotation operator for the $M$ (occupied + virtual) orbitals with $\hat{\kappa}(\bm{\kappa}) = \sum_{k=1}^M \sum_{l=1}^M  \kappa_{kl} \,  \hat{E}_{kl}$ where $\kappa_{kl}$ are the orbital rotation parameters and $\hat{E}_{kl}$ is the singlet excitation operator from orbital $l$ to orbital $k$, $\hat{E}_{kl}=\hat{a}_{k \uparrow}^{\dag} \hat{b}_{l \uparrow} + \hat{a}_{k \downarrow}^\dag \hat{b}_{l
\downarrow}$, written with dual biorthogonal orbital creation and annihilation operators $\hat{a}_{k \sigma}^\dag$ and
$\hat{b}_{l \sigma}$ (see, e.g., Ref.~\onlinecite{HelJorOls-BOOK-02}). The parameters $\b{p}=(\bm{\alpha},\b{c},\bm{\kappa})$ to be optimized are the parameters in the Jastrow factor $\bm{\alpha}$, the VB structure coefficients $\b{c}$, and the orbital rotation parameters $\bm{\kappa}$.

For orbital optimization, the orbitals are partitioned into three sets: inactive (doubly occupied in all determinants), active
(occupied in some determinants and unoccupied in others), and virtual (unoccupied in all determinants). All inactive and active orbitals are optimized. The non-redundant excitations to be considered are inactive $\to$ active, inactive $\to$ virtual, active $\to$ virtual and active $\to$ active. If the
action of the excitation $\hat{E}_{kl}$ on the wave function is not zero but the reverse excitation $\hat{E}_{lk}$ is zero, then
the orthogonality condition $\kappa_{lk} = - \kappa_{kl}$ is imposed. For some active-active excitations, both direct and
reverse excitations ($\hat{E}_{kl}$ and $\hat{E}_{lk}$) may be allowed, and thus it makes sense for localized orbitals to
remove the orthogonality constraint by treating $\kappa_{kl}$ and $\kappa_{lk}$ as independent parameters. This results in
only very few (if any at all) additional orbital parameters for the wave functions considered here. We note that, when
considering active-active excitations, redundancies between two orbital wave-function derivatives or between an orbital
wave-function derivative and a VB structure frequently occur, and must be detected and eliminated. Localized orbitals do not have the point
group symmetry of the system, so the number of orbital excitations cannot be reduced based on the non mixing of irreducible
representations, as usually done. However, the number of orbital excitations is greatly reduced by forbidding mixing between
orbitals of different localization classes, i.e. expanded on different subsets of basis functions. 

When a single set of orbitals is used for all VB structures, the wave function in Eq.~(\ref{PsiJVBp}) corresponds to a Valence-Bond self-consistent field (VBSCF)~\cite{LenBal-JCP-83} wave function multiplied by a Jastrow factor. It is easy to extend Eq.~(\ref{PsiJVBp}) to the case of a BOVB expansion in which each VB structure is allowed to have a different set of orbitals, which can be written as
\begin{eqnarray}
\ket{\Psi(\b{p})} = \hat{J}(\bm{\alpha}) \, \sum_{I=1}^{N_\text{VB}} c_I \, e^{\hat{\kappa}_I(\bm{\kappa}_I)} \ket{\Phi_I},
\label{PsiJBOVB}
\end{eqnarray}
where $\hat{\kappa}_I(\bm{\kappa}_I)$ is the orbital operator for the set of orbitals in the VB structure $\ket{\Phi_I}$. In this case, only (occupied and unoccupied) orbitals that belong to the same set are allowed to mix in the optimization. 

The $N_\p$ parameters $\b{p}$ are optimized by minimizing the energy using the linear optimization method~\cite{TouUmr-JCP-07,UmrTouFilSorHen-PRL-07,TouUmr-JCP-08}. The idea of the method is to iteratively:\\
(i) expand the normalized wave function
$\ket{\Psib(\b{p})}=\ket{\Psi(\b{p})}/\sqrt{\braket{\Psi(\b{p})}{\Psi(\b{p})}}$
to first order in the parameter variations $\Delta \b{p} = \b{p} - \b{p}^0$ around the current
parameters $\b{p}^0$
\begin{eqnarray}
\ket{\Psib_\lin(\b{p})} = \ket{\Psib_0} + \sum_{j=1}^{N_\p} \Delta p_j \, \ket{\Psib_j},
\label{Psiblin}
\end{eqnarray}
where $\ket{\Psib_0} = \ket{\Psib(\b{p}^0)}$ is the normalized current wave function and $\ket{\Psib_j}$ are the first-order derivative of the normalized wave function with respect to the parameters at $\b{p}^0$,
\begin{eqnarray}
\ket{\Psib_j}=\left.\frac{\partial \ket{\Psib(\b{p})}}{\partial p_j}\right|_{\b{p}=\b{p}^0} = \frac{1}{\sqrt{\braket{\Psi_0}{\Psi_0}}} \left(\ket{\Psi_j} - \frac{\braket{\Psi_0}{\Psi_j}} {\braket{\Psi_0}{\Psi_0}} \ket{\Psi_0} \right),
\label{semiorthog}
\end{eqnarray}
written in terms of the unnormalized current wave function $\ket{\Psi_0} = \ket{\Psi(\b{p}^0)}$ and its first-order derivatives $\ket{\Psi_j}=\partial \ket{\Psi(\b{p})}/\partial p_j|_{\b{p}=\b{p}^0}$;\\
(ii) minimize the expectation value of the
Hamiltonian $\hat{H}$ over this linear wave function with respect to the parameter variations $\Delta \b{p}$
\begin{eqnarray}
E_\lin = \min_{\Delta \b{p}} \frac{\bra{\Psib_\lin(\b{p})} \hat{H} \ket{\Psib_\lin(\b{p})}}{\braket{\Psib_\lin(\b{p})}{\Psib_\lin(\b{p})}};
\end{eqnarray}
(iii) update the current parameters as $\b{p}^0 \to \b{p}^0 +  \Delta \b{p}$.

The energy minimization step (ii) is equivalent to finding the lowest solution of the $(N_\p+1)$-dimensional generalized eigenvalue equation
\begin{eqnarray}
 \left(\begin{array}{cc} E_0 & \b{g}^\T/2\\
                              \b{g}/2 & \overline{\b{H}}\\
      \end{array} \right) \left( \begin{array}{c}    1 \\ \Delta \b{p} \\ \end{array} \right)
= E_\lin
 \left( \begin{array}{cc}    1 & \b{0}^\T\\
                            \b{0} & \overline{\b{S}}\\
      \end{array} \right) \left( \begin{array}{c}    1 \\ \Delta \b{p} \\ \end{array} \right),
\label{geneigeq}
\end{eqnarray}
where $E_0= \bra{\Psib_0} \hat{H} \ket{\Psib_0}$ is the current energy, $\b{g}$ is the gradient of the energy with respect to the $N_\p$ parameters
with components $g_i=2 \bra{\Psib_i} \hat{H} \ket{\Psib_0}$, $\overline{\b{H}}$ is the Hamiltonian matrix in the basis consisting of the $N_\p$
wave function derivatives with elements $\overline{H}_{ij}=\bra{\Psib_i} \hat{H} \ket{\Psib_j}$, and $\overline{\b{S}}$ is the overlap matrix
in this basis with elements $\overline{S}_{ij}=\braket{\Psib_i}{\Psib_j}$.

\subsection{Quantum Monte Carlo implementation}

\subsubsection{Variational Monte Carlo}

The linear optimization method is realized in VMC. The idea of the VMC method~\cite{Mil-PR-65,CepCheKal-PRB-77} (see Ref.~\onlinecite{TouAssUmr-INC-16} for a review) is simply to calculate the multidimensional integrals appearing in quantum mechanics using a Monte Carlo numerical integration technique. For example, the energy $E_0$ of the current wave function $\Psi_0$ is written as
\begin{equation}
E_0 = \frac{\bra{\Psi_0} \hat{H} \ket{\Psi_0}}{\braket{\Psi_0}{\Psi_0}} = \int_{\mathbb{R}^{3N}} \d\b{R} \, E_\L(\b{R}) \, \rho_0(\b{R}),
\end{equation}
where $E_\L(\b{R}) = (H\Psi_0(\b{R}))/\Psi_0(\b{R})$ is the local energy depending on the $3N$ spatial electron coordinates $\b{R}=(\b{r}_1,\b{r}_2,...,\b{r}_N)$, and $\rho_0(\b{R})= \Psi_0(\b{R})^2/\int \d\b{R} \Psi_0(\b{R})^2$ is the (normalized) probability density. Note that, as usual in QMC, the spin coordinates have been fixed~\cite{HuaFilUmr-JCP-98}, so that we consider functions of the spatial coordinates $\b{R}$ only. The energy $E_0$ can then be estimated as the average value of $E_\L(\b{R})$ on a sample of $M$ points $\b{R}_k$ sampled from the probability density $\rho_0(\b{R})$,
\begin{equation}
E_0 \approx {^M}E_0 = \langle E_\L(\b{R}) \rangle_{\rho_0} = \frac{1}{M} \sum_{k=1}^{M} E_\L(\b{R}_k),
\label{Elav}
\end{equation}
where we have introduced the compact notation ${^M}E_0$ to designate the average over the sample of $M$ points. In practice, the points $\b{R}_k$ are sampled using a random walk following the Metropolis-Hastings algorithm~\cite{MetRosRosTelTel-JCP-53,Has-B-70}. The advantage of this approach is that it does not use an analytical integration involving the wave function, and thus does not impose severe constraints on the form of the wave function.

Similarly, the VMC version of the generalized eigenvalue equation of Eq.~(\ref{geneigeq}) is~\cite{NigMel-PRL-01,TouUmr-JCP-07,UmrTouFilSorHen-PRL-07,TouUmr-JCP-08}
\begin{eqnarray}
 \left(\begin{array}{cc} {^M}E_0 & ^M\b{g}_\R^\T/2\\
                              ^M\b{g}_\L/2 & ^M\overline{\b{H}}\\
      \end{array} \right) \left( \begin{array}{c}    1 \\ \Delta \b{p} \\ \end{array} \right)
= E_\lin
 \left( \begin{array}{cc}    1 & \b{0}^\T\\
                            \b{0} & ^M\overline{\b{S}}\\
      \end{array} \right) \left( \begin{array}{c}    1 \\ \Delta \b{p} \\ \end{array} \right),
\label{geneigeqM}
\end{eqnarray}
where $^M\b{g}_\L$ and $^M\b{g}_\R$ are two estimates of the energy gradient with components
\begin{eqnarray}
^Mg_{\L,i} &=& 2 \left\langle \frac{\Psib_i (\b{R})}{\Psib_0(\b{R})} \frac{H \Psib_0(\b{R})}{\Psib_0(\b{R})} \right\rangle_{\!\!\rho_0} \nonumber \\
&=& 2 \Biggl[ \left\langle \frac{\Psi_i(\b{R})}{\Psi_0(\b{R})} E_\L(\b{R}) \right\rangle_{\!\!\rho_0} - \left\langle \frac{\Psi_i(\b{R})}{\Psi_0(\b{R})}
\right\rangle_{\!\!\rho_0} \left\langle E_\L(\b{R}) \right\rangle_{\rho_0} \Biggl],
\nonumber\\
\label{Hi0}
\end{eqnarray}
where $\Psib_i (\b{R})/\Psib_0(\b{R}) = \Psi_i (\b{R})/\Psi_0(\b{R}) - \left\langle \Psi_i (\b{R})/\Psi_0(\b{R}) \right\rangle$ has been used,
\begin{eqnarray}
^Mg_{\R,j} &=& 2 \left\langle \frac{H \Psib_j(\b{R})}{\Psib_0(\b{R})} \right\rangle_{\!\!\rho_0} \nonumber \\
&=& 2 \Biggl[ \left\langle \frac{\Psi_j(\b{R})}{\Psi_0(\b{R})} E_\L(\b{R}) \right\rangle_{\!\!\rho_0} - \left\langle \frac{\Psi_j(\b{R})}{\Psi_0(\b{R})}
\right\rangle_{\!\!\rho_0} \left\langle E_\L(\b{R}) \right\rangle_{\rho_0}
+ \left\langle E_{\L,j}(\b{R}) \right\rangle_{\rho_0} \Biggl],
\label{H0j}
\end{eqnarray}
where $E_{\L,j}(\b{R}) = (H \Psi_j(\b{R}))/\Psi_0(\b{R}) - \left[ \Psi_j(\b{R})/\Psi_0(\b{R}) \right] E_{\L}(\b{R})$ is the
derivative of the local energy with respect to the parameter $p_j$ (whose average is zero in the limit of an infinite sample),
$^M\overline{\b{H}}$ is the following nonsymmetric estimate
of the Hamiltonian matrix
\begin{eqnarray}
^M\overline{H}_{ij} &=& \left\langle \frac{\Psib_i (\b{R})}{\Psib_0(\b{R})} \frac{H \Psib_j(\b{R})}{\Psib_0(\b{R})} \right\rangle_{\!\!\rho_0}
\nonumber\\
&=&\left\langle \frac{\Psi_i(\b{R})}{\Psi_0(\b{R})} \frac{\Psi_j(\b{R})}{\Psi_0(\b{R})} E_\L(\b{R}) \right\rangle_{\!\!\rho_0}
- \left\langle \frac{\Psi_i(\b{R})}{\Psi_0(\b{R})} \right\rangle_{\!\!\rho_0}  \left\langle \frac{\Psi_j(\b{R})}{\Psi_0(\b{R})} E_\L(\b{R}) \right\rangle_{\!\!\rho_0}
\nonumber\\
&&- \left\langle \frac{\Psi_j(\b{R})}{\Psi_0(\b{R})} \right\rangle_{\!\!\rho_0}  \left\langle \frac{\Psi_i(\b{R})}{\Psi_0(\b{R})} E_\L(\b{R}) \right\rangle_{\!\!\rho_0}
+ \left\langle \frac{\Psi_i(\b{R})}{\Psi_0(\b{R})} \right\rangle_{\!\!\rho_0}  \left\langle \frac{\Psi_j(\b{R})}{\Psi_0(\b{R})} \right\rangle_{\!\!\rho_0}  \left\langle
E_\L(\b{R}) \right\rangle_{\rho_0}
\nonumber\\
&&
+ \left\langle \frac{\Psi_i(\b{R})}{\Psi_0(\b{R})} E_{\L,j}(\b{R}) \right\rangle_{\!\!\rho_0} - \left\langle \frac{\Psi_i(\b{R})}{\Psi_0(\b{R})}
\right\rangle_{\!\!\rho_0} \left\langle E_{\L,j}(\b{R}) \right\rangle_{\rho_0},
\label{Hij}
\end{eqnarray}
and $^M\overline{\b{S}}$ is the estimated overlap matrix
\begin{eqnarray}
^M\overline{S}_{ij}&=& \left\langle \frac{\Psib_i (\b{R})}{\Psib_0(\b{R})} \frac{\Psib_j(\b{R})}{\Psib_0(\b{R})} \right\rangle_{\!\!\rho_0}
\nonumber\\
&=&\left\langle \frac{\Psi_i(\b{R})}{\Psi_0(\b{R})} \frac{\Psi_j(\b{R})}{\Psi_0(\b{R})} \right\rangle_{\!\!\rho_0}  - \left\langle
\frac{\Psi_i(\b{R})}{\Psi_0(\b{R})} \right\rangle_{\!\!\rho_0} \left\langle \frac{\Psi_j(\b{R})}{\Psi_0(\b{R})} \right\rangle_{\!\!\rho_0}.
\label{Sij}
\end{eqnarray}
Using these non-symmetric estimators in Eqs.~(\ref{Hi0}),~(\ref{H0j}),~(\ref{Hij}) ensures a strong zero-variance principle~\cite{NigMel-PRL-01}: In the limit where the current wave function and its first-order derivatives with respect to the parameters form a complete basis of the Hilbert space considered, the optimal parameters variations $\Delta \b{p}$ are obtained from Eq.~(\ref{geneigeqM}) with zero variance. In practice, we are never in this limit of course, but nevertheless solving the generalized eigenvalue equation of Eq.~(\ref{geneigeqM}) leads to parameter variations with significantly smaller statistical fluctuations than the parameter variations that would be obtained with a symmetrized version of the generalized eigenvalue equation. Of course, in the limit of an infinite sample $M \to \infty$, the generalized eigenvalue equation of Eq.~(\ref{geneigeqM}) properly reduces to the symmetric generalized eigenvalue equation of Eq.~(\ref{geneigeq}).

\subsubsection{Diffusion Monte Carlo}

Once the JVB wave function $\Psi_0$ with optimal parameters has been obtained in VMC, a more accurate wave function can be calculated using DMC~\cite{GriSto-JCP-71,And-JCP-75,And-JCP-76,ReyCepAldLes-JCP-82,MosSchLeeKal-JCP-82} (see Ref.~\onlinecite{TouAssUmr-INC-16} for a review). Namely, in the FN approximation~\cite{And-JCP-75,KlePic-JCP-76,And-JCP-76}, one obtain the FN wave function $\Psi_\FN$ by applying to $\Psi_0$ the imaginary-time propagator in the long-time limit
\begin{equation}
\ket{\Psi_\FN} \propto \lim_{t\to\infty} e^{-(\hat{H}_\FN-E_\T)t} \ket{\Psi_0},
\end{equation}
where $\hat{H}_\FN$ is the FN Hamiltonian, which can be thought of as obtained by adding to the true Hamiltonian $\hat{H}$ infinite potential barriers at the location of the nodes of $\Psi_0$~\cite{BadHayNee-PRB-08}, and $E_\T$ is a trial energy adjusted to properly converge to the lowest-energy state. The obtained FN wave function $\Psi_\FN$ is the best variational wave function having the same nodes as the wave function $\Psi_0$. The corresponding FN energy can be written as (see, e.g., Ref.~\onlinecite{TouAssUmr-INC-16})
\begin{equation}
E_\FN = \frac{\bra{\Psi_\FN} \hat{H} \ket{\Psi_0}}{\braket{\Psi_\FN}{\Psi_0}} = \int_{\mathbb{R}^{3N}} \d\b{R} \, E_\L(\b{R}) \, \rho_\FN(\b{R}),
\end{equation}
where $E_\L(\b{R}) = (H\Psi_0(\b{R}))/\Psi_0(\b{R})$ is the same local energy as in VMC and $\rho_\FN(\b{R}) = \Psi_\FN(\b{R}) \Psi_0(\b{R})/ \int \d\b{R} \Psi_\FN(\b{R}) \Psi_0(\b{R})$ is the mixed FN (normalized) probability density. 

In practice, $\rho_\FN(\b{R})$ is sampled using a weighted random walk involving at each step $k$ a population of $M_k$ walkers (i.e., points) $\b{R}_{k,m}$ undergoing a birth/death process to control the fluctuations of their weights $w_{k,m}$. The FN energy can then be estimated as the weighted average of the local energy over the $M$ steps and $M_k$ walkers
\begin{equation}
E_\FN \approx \langle E_\L(\b{R}) \rangle_{\rho_\FN} = \frac{\sum_{k=1}^{M} \sum_{m=1}^{M_k} w_{k,m} E_\L(\b{R}_{k,m})}{\sum_{k=1}^{M} \sum_{m=1}^{M_k} w_{k,m}}.
\label{E0DMCweights}
\end{equation}

\subsection{Weights of the Jastrow-Valence-Bond structures}

One of the great interests of VB wave functions are their unique interpretative capabilities, as will be illustrated in the examples of Sections~\ref{sec:benchmarking} and~\ref{sec:applications}, and as illustrated in many other chapters of this book. This can be traced back to the direct mapping between the mathematical expressions of the VB structures and the pictorial Lewis model. As such, the most commonly used interpretative quantities that directly come out from optimized VB wave functions are probably the weights of the VB structures, which provide to chemists a quantitative weighted picture of the different structures in a given molecule and a given electronic state. It is therefore of utmost importance to retain the interpretative power of the more classical VB methods that the implementation of the VB-QMC method allows one to compute VB weights. Note that weights are presently only available at the VB-VMC level.

The JVB wave functions in Eq.~(\ref{PsiJVBSCF}) can be rewritten as
\begin{eqnarray}
\ket{\Psi_0} = \sum_{I=1}^{N_\text{VB}} c_I \ket{\Psi_I},
\label{}
\end{eqnarray}
where $\ket{\Psi_I} = \hat{J} \ket{\Phi_I}$ are products of the Jastrow-factor operator $\hat{J}$ and VB structures $\ket{\Phi_I}$. The JVB structures $\{ \ket{\Psi_I} \}$ are non-orthogonal. Similarly to standard VB, we can define different weights associated with the JVB structures. We will consider the two main definitions: the Chirgwin-Coulson weights and the L\"owdin weights.

\begin{itemize}
\item The Chirgwin-Coulson weights~\cite{ChiCou-PRSL-50} (also called Mulliken weights) are based on expanding the squared norm of the wave function as
\begin{eqnarray}
\braket{\Psi_0}{\Psi_0} =\sum_{I=1}^{N_\text{VB}} \sum_{J=1}^{N_\text{VB}} c_I c_J S_{I,J},
\label{}
\end{eqnarray}
where $S_{I,J} = \braket{\Psi_I}{\Psi_J}$ is the overlap matrix of the JVB structures. This naturally leads to the definition of the Chirgwin-Coulson (CC) weights
\begin{eqnarray}
w_I^\text{CC} = \frac{\sum_{J=1}^{N_\text{VB}} c_I c_J S_{I,J}}{\braket{\Psi_0}{\Psi_0}},
\label{}
\end{eqnarray}
which properly sum to unity: $\sum_{I=1}^{N_\text{VB}} w_I^\text{CC} =1$.

\item The L\"owdin weights are based on L\"owdin's symmetric orthogonalization~\cite{Low-AQC-70} of the JVB structures which permits to rewrite the wave function as
\begin{eqnarray}
\ket{\Psi_0} = \sum_{K=1}^{N_\text{VB}} \bar{c}_K \ket{\bar{\Psi}_K},
\label{}
\end{eqnarray}
where $\ket{\bar{\Psi}_K}= \sum_{I=1}^{N_\text{VB}} (S^{-1/2})_{K,I} \ket{\Psi_I}$ are the symmetrically orthonormalized JVB structures and $\bar{c}_K = \sum_{I=1}^{N_\text{VB}} c_I (S^{1/2})_{I,K}$ are the associated coefficients. This leads to the definition of the L\"owdin (L) weights
\begin{eqnarray}
w_K^\text{L} = \frac{\bar{c}_K^2}{\braket{\Psi_0}{\Psi_0}},
\label{}
\end{eqnarray}
where $\braket{\Psi_0}{\Psi_0}= \sum_{K=1}^{N_\text{VB}} \bar{c}_K^2$. Again, the weights properly sum to unity: $\sum_{K=1}^{N_\text{VB}} w_K^\text{L} =1$.
\end{itemize}

Thus, the computation of the weights only requires the overlap matrix $S_{I,J}$ which is estimated in VMC by
\begin{eqnarray}
S_{I,J} = \left\langle \frac{\Psi_I(\b{R})}{\Psi_0(\b{R})} \frac{\Psi_J(\b{R})}{\Psi_0(\b{R})} \right\rangle_{\!\!\rho_0},
\label{}
\end{eqnarray}
which was already needed for the optimization of the coefficients $\{c_I\}$ of the VB structures [see Eq.~(\ref{Sij})].

\subsection{Software implementation: CHAMP}

The VB-QMC method is implemented in the software CHAMP~\cite{Cha-PROG-XX}. All-electron and pseudopotential calculations are available.
Initial VB wave functions can be read from the software XMVB~\cite{SonMoZhaWu-JCC-05}. The parameters in the JVB wave functions are then simultaneously optimized with the linear optimization method in VMC, using an accelerated
Metropolis algorithm~\cite{Umr-PRL-93,Umr-INC-99}. Once the trial wave function has been optimized, a DMC calculation can be performed within the short-time and FN approximations, using an efficient DMC algorithm featuring very small time-step errors~\cite{UmrNigRun-JCP-93}. Typically, we use an imaginary time step of $\tau=0.01$ Hartree$^{-1}$.

\section{Benchmarking the Valence-Bond Quantum Monte Carlo method}
\label{sec:benchmarking}

\subsection{Proof of concept: Bond dissociation energy of acetylene}

One of the first published studies to use ab initio VB trial wave functions in QMC was the study of the C-H bond dissociation of acetylene~\cite{DomBraLes-JPCA-08}. While there were earlier QMC studies that used geminals~\cite{CasSor-JCP-03} and resonating VB wave functions\cite{CasAttSor-JCP-04}, the study of the bond dissociation of acetylene in Ref.~\onlinecite{DomBraLes-JPCA-08} examined the use of chemically interpretable VB trial functions, namely BOVB, in QMC for calculating molecular bond dissociation energies. It was already well known that the accuracy of QMC results could be improved by using multideterminant trial wave functions~\cite{FlaCafSav-INC-97} rather than single-determinant trial wave functions, unfortunately the commonly used complete-active-space self-consistent-field (CASSCF) and configuration-interaction (CI) expansions usually become unwieldy large for molecules containing more than a few atoms. The localized nature and non-orthogonality of VB orbitals usually allow for wave functions with significantly fewer Slater determinants than CASSCF or CI expansions while providing easy to interpret description of the qualitative nature of the covalent bonds present within the molecules.

The equilibrium molecular geometries and scaled harmonic zero-point vibrational energies were calculated with density-functional theory using the B3LYP functional with the cc-pVTZ basis set, rather than using experimental, VB, or QMC geometries. This common practice allows one to economically treat molecular systems larger than diatomics or triatomics while introducing an error that is expected to be less than 1-2 kcal/mol, which is roughly the order of magnitude of the experimental uncertainty of the homolytic bond dissociation energy. In this pathfinder study only the parameters of a Schmidt-Moskowitz-Boys-Handy Jastrow factor were optimized, and the determinantal parameters of the wave functions were kept unchanged from the VB calculations. This practical choice was made to substantially reduce the computational effort needed to optimize the JVB trial functions, full optimization of Jastrow-Slater multideterminant wave functions at the QMC level being still a challenge at the time of this study, and was rationalized by the observation that DMC with a simple trial wave function in general recovers more of the electronic correlation energy than VMC with a more elaborate trial wave function. Furthermore, the hypothesis of the study was that given that single-determinant mean-field trial wave functions already have nodal surfaces that recover more than 90\% of the correlation energy, multiconfiguration mean-field trial functions that are complete within their active space should recover a substantial percentage of the missing correlation energy and that the remaining missing correlation energy should be small in comparison. Besides, because BOVB wave functions are compact wave functions that include both static and dynamical correlation, it was expected that BOVB nodes might lead to lower subsequent DMC energy compared to the DMC energy obtained with nodes from wave functions that include static correlation only (VBSCF or CASSCF).  The level of VB theory used in the construction of the JVB trial wave functions was ``partial BOVB'' (p-BOVB) in which different VB structures share the same set of orbitals rather than letting the orbitals fully relax in each VB structure which would be the full split-delocalized BOVB (SD-BOVB) description. The VB active space involved the carbon-carbon $\pi$ and $\sigma$ bonds as well as the carbon-hydrogen  $\sigma$ bond that is broken during the dissociation. This treatment resulted in 7 VB structures (14 Slater determinants) for the ethynyl radical (C$_2$H) and 21 VB structures (56 Slater determinants) for the acetylene molecule (C$_2$H$_2$).   

\begin{center}
\begin{table}
\caption{Estimated percentage of the electronic correlation energy (compared to estimated exact total electronic energies from Ref. ~\onlinecite{doi:10.1063/1.1335596}) for C$_2$H and C$_2$H$_2$, and the C-H bond dissociation energy (BDE) for C$_2$H$_2$ (in kcal/mol) calculated with several methods. For all methods, the TZP basis set is used.}
\label{table:DTBE}
\begin{tabular}{|l|c|c|c|c|c|}
\hline     
  &  p-BOVB & CCSD(T) & HF-DMC$^a$ &  p-BOVB-DMC$^b$ & Exp.\\
\hline
C$_2$H & 18.6\% & 60.3\% &  94.3(1)\%     & 95.3(2)\%    & \\  
C$_2$H$_2$  & 19.9\% & 61.4\% & 96.2(1)\%       & 95.4(3)\%    & \\
\hline
BDE    & 120.0 & 128.2 &  137.5(5)         & 132.4(9)   &  132.8(7)$^c$ \\
\hline
\end{tabular}\\
$^a$DMC with a non-reoptimized HF wave function multiplied by a Jastrow factor.\\[-0.2cm]
$^b$DMC with a non-reoptimized p-BOVB wave function multiplied by a Jastrow factor.\\[-0.2cm]
$^c$Recommended experimental BDE from Ref.~\onlinecite{HandBook}.
\end{table}
\end{center}
In Table \ref{table:DTBE} we see the percentage of the electronic correlation energy recovered for C$_2$H and C$_2$H$_2$ along with the resulting C-H bond dissociation energy (BDE).  In the first column we see that on its own, without a Jastrow factor, the p-BOVB method using a triple-zeta Slater basis set (TZP) only recovers 18.6-19.9\% of the correlation energy which results in a BDE that is off by almost 13 kcal/mol. In comparison, for this basis set, the ``gold standard'' CCSD(T) method recovers 60.3-61.4\% of the correlation energy and only rises to 72.7-74.9\% if one uses a larger quadruple-zeta basis set (not shown). In contrast, DMC with a simple single-determinant trial wave function based on spin-restricted Hartree-Fock (HF) orbitals gives 94.3-96.2\% of the correlation energy and its BDE is in error of only about 5 kcal/mol.  With a Jastrow-Slater p-BOVB trial wave function, the DMC BDE is $132 \pm 0.9$ kcal/mol, which very well matches the experimental value of $132.8 \pm 0.7$ kcal/mol. This is due to the fact that for both the acetylene and the ethynyl radical the percentage of the recovered electronic correlation energy is the same, which was expected thanks to the well-balanced treatment of the electronic correlation which is a prominent feature of the BOVB method (see corresponding chapter in this book). Note that this consistency for the recovered correlation energy of a hydrocarbon and its bond dissociation radical product was also previously seen for other types of trial wave functions with the DMC method \cite{ doi:10.1002/kin.20063}. Clearly, at the DMC level, using p-BOVB trial wave functions, instead of a HF single-determinant wave function, improves the description of the C$_2$H radical.

However, DMC with the p-BOVB trial wave function only recovers a comparable amount of correlation energy than DMC with the HF single-determinant trial wave function for C$_2$H$_2$. Therefore, this study  partially invalidated the hypothesis that reoptimization of the parameters of determinantal part of the JVB would not be essential in practice for improving DMC energies, and hence rather highlights the importance of optimizing all the parameters of trial function when benchmark DMC calculations are necessary. As consequence, this preliminary study did not manage to convincingly illustrate the potential merits of using BOVB-type trial wave functions in DMC as compared with standard single or multi-determinant expansions based on delocalized orbitals. In what follows, studies are presented that make use of fully optimized JVB wave functions.

\subsection{Total energies of a selection of small molecules}

Building upon the knowledge learned from the acetylene bond dissociation study, a follow-up VB-QMC study~\cite{BraTouCafUmr-JCP-11} examined the total energies and well depths of 4 first-row homonuclear diatomic molecules (C$_2$, N$_2$, O$_2$, and F$_2$) at their experimental bond lengths. Unlike the previous work, Slater determinant coefficients, orbitals, and basis exponents were optimized simultaneously with the Jastrow parameters in VMC. The work in Ref.~\onlinecite{BraTouCafUmr-JCP-11} also explored the differential impact of optimizing these various wave function parameters in both single bonding-pattern and multiple bonding-pattern VB wave functions and compared them to full-valence complete-active-space (CAS) Jastrow-Slater wave functions.

The concept of "bonding pattern" used here denotes a specific coupling of the valence electrons of a molecule into pairs, and at the same time the complete set of classical VB structures associated with it. A bonding pattern is therefore associated with a specific fully covalent structure, and to a given chemical (Lewis) structure, but it also denotes the group of classical (covalent and ionic) VB structures necessary to fully account for the static (left-right) correlation associated with each bond.
The ground state of the F$_2$ molecule ($^1\Sigma_g^+$) can be well represented by a single bonding-pattern, (i.e., a $\sigma$-bond between the F atoms) and therefore adding only a few additional bonding-pattern configurations does not substantially change neither the calculated  VBSCF nor QMC well-depths. In fact, F$_2$ along with O$_2$ have substantially more dynamic electronic correlation than C$_2$ or N$_2$, thus adding additional bonding-patterns primarily serves to recover some of the missing dynamic correlation rather than static correlation. The Jastrow factor is by far more efficient in recovering dynamic correlation than adding more configuration state functions or VB structures, hence it is not surprising that for species with mono-configurational character (i.e. both small static correlation and large dynamic correlation effects) single bonding-pattern wave functions tend to be sufficient for  most QMC calculations. Due to the high symmetry of the F$_2$ molecule, the multiple bonding-pattern VBSCF wave function has as many determinants as full-valence CAS, and in fact requires mixing more excited state determinants during the super-configuration-interaction-like wave-function optimization than the full-valence CAS wave function. This situation is not seen in the other diatomics in the study, and tends to be less of an issue for systems with multiple bonds and less symmetric systems, where localized orbitals should generally give a more compact representation than canonical delocalized orbitals. 
The diatomic molecules C$_2$ ($^1\Sigma_g^+$), N$_2$ ($^1\Sigma_g^+$), and O$_2$ ($^3\Sigma_g^-$) are best described by single bonding-patterns consisting of their major Lewis structures. C$_2$ and N$_2$ displays triple bonds ($\sigma$ bond with 2 $\pi$ bonds) with either a weaker fourth singlet-coupling for C$_2$ that may be considered as a fourth bond, or two lone pairs on opposite atoms for N$_2$. The dioxygen molecule in its triplet ground state displays a 2-electron $\sigma$ bond and two $\pi$ 3-electron half bonds, with a lone pair of electrons on each oxygen atom. 

\begin{center}
\begin{table}
\caption{Estimated percentage of the total electronic correlation energy for first-row diatomics recovered with VMC and DMC with different trial wave functions. The VB and BOVB wave functions correspond to single-bonding patterns. For all cases, the coefficients of the determinants and the orbitals have been optimized together with the Jastrow factor in VMC. The core-valence triple-zeta quality Slater basis set (CVB1) of Ema \textit{et al.}~\cite{EmaGarRamLopFerMeiPal-JCC-03} is used.}
\label{table:DCOR}
\begin{tabular}{|l|c|c|c|c|}
\hline     
        &  C$_2$ & N$_2$ & O$_2$ & F$_2$ \\
\hline
HF-VMC & 78.4\% & 83.6\% & 85.3\%& 85.7\% \\
VB-VMC & 88.0\% & 87.4\% & 87.2\% & 88.3\% \\
BOVB-VMC & - & - & 88.4\% & 89.0\% \\
HF-DMC & 88.6\%& 93.0\%& 93.8\% & 93.9\%\\
VB-DMC & 94.1\%& 94.5\%& 94.3\% & 94.7\% \\
BOVB-DMC & - & - & 94.8\% & 95.0\% \\
CAS-DMC$^a$ & 97.0\%& 96.0\%& 95.1\%& 95.7\% \\
\hline
\end{tabular}\\
$^a$DMC with full-valence CAS wave function from Ref.~\onlinecite{TouUmr-JCP-08}.
\end{table}
\end{center}

In Table~\ref{table:DCOR} are estimated the percentage of the total electronic correlation energy various methods can recover. The VB and BOVB wave functions correspond to single-bonding patterns. For C$_2$, a single-determinant wave function poorly describes the main bonding-pattern for this diatomic, mainly because of an inherent constraint in the weight of multiionic structures of very different stability (for instance, within a single-determinantal description, the C$^{4+}$ C$^{4-}$ fully ionic structure will have the same weight as neutral $^{2-}$C$^{2+}$ $^{2-}$C$^{2+}$ structure). Therefore, there is a significant improvement when a single bonding-pattern VBSCF-type wave function is used to described the molecule. At the VMC level, the correlated HF-VMC calculation recovers 78.4\% of the correlation energy while VB-VMC recovers 88.0\%. This improvement is seen also at the DMC level with VB-DMC recovering 5.5\% more correlation energy than HF-DMC. Singlet diradicals need at least two configurations to be qualitatively described at the mean-field level. For the remaining diatomics, a single-determinant wave function provides a more reasonable description, thus HF-VMC and HF-DMC energies are closer to VB-VMC and VB-DMC. Finally, it is interesting to note that going to a BOVB description does not substantially improve the VMC and DMC results.

\subsection{Weights and resonance energies}

In this section we provide examples of the chemical insight that can be obtained using the VB-QMC method. It will be demonstrated that the VB-QMC method can provide very accurate electronic energies and energy differences on one side, and on the other side the obtained wave functions preserve all interpretative capabilities captured into the parent VB wave functions. In particular, it will be shown how the proposed VB-QMC approach can be successfully applied to the calculations of resonance energies and weights of VB structures. The concept of resonance is key in chemistry, and relates to the case where a molecule cannot be described by a single Lewis structure.~\cite{paulingbook} The resonance energy is the energy difference, which can be computed within VB theory, between a ``diabatic" state and the ``adiabatic" state. A ``diabatic" state is an electronic state that maps at any geometry to a given set of Lewis structures. This mapping is often to the most stable Lewis structure. The ``adiabatic" state corresponds to a mixing of several Lewis structures.

The geometries of the studied molecules were optimized with M{\o}ller-Plesset second-order (MP2) perturbation theory. These calculations were performed using the energy-consistent effective core potentials with the corresponding VTZ basis set of Burkatzki {\it et al.}~\cite{Burkatzki-Basis-JCP-2007}. The CCSD(T) single-point energy calculations were used to obtain the reference energy values. In VB, CCSD(T), and VB-QMC calculations we employed the VTZ basis set from which f basis functions were removed. 

\subsubsection{Charge-shift resonance energies of some homonuclear bonds}
\label{sec:csreso}

A new type of chemical bonding, the so-called ``charge-shift bonding'' (CSB), emerged from a series of systematic VB studies (see Refs.~\onlinecite{ShaDanGalBraWuHib-ACIE-20,ShaDanGalBraWuHib-AC-20}). CSB appears as a third distinct class of bonding along with the traditional covalent and ionic bonding mechanisms, where bonding does not arise from the spin-pairing of the covalent structure or from the electrostatic stabilization attached to the ionic structure, but predominantly from the resonance energy arising from the covalent-ionic mixing, and sometimes plays a decisive role in some unusual bonding situations. The usual characterization of the bonding type within classical VB theory is obtained by calculating the charge-shift resonance energy (RE$_\text{CS}$), which is, for a two-center two-electron bond, the energy difference between the most stable covalent or ionic structure and the full multi-structure adiabatic state.  When RE$_\text{CS}$ exceeds 50\% of the total bond dissociation energy the bond is then qualified as a charge-shift bond (and when it exceeds 100\% it is sometimes called a ``complete charge-shift bond'').

\begin{figure}[t]
    \centering
    \includegraphics[width=15cm]{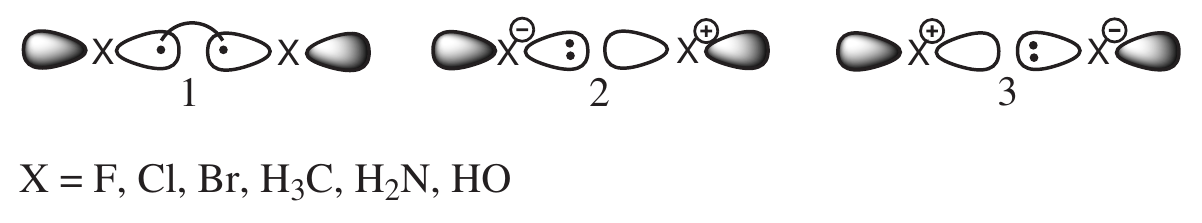}
    \caption{Set of VB structures for the studied molecules given in Table~\ref{tab:dissociation}.}
    \label{fig:VBstructures}
\end{figure}

The studied series of molecules is given in Table~\ref{tab:dissociation}. In a recent work~\cite{Shaik-CSB-NaturChem-2009}, it has been shown that in this series of molecules different types of bonding can be found with regard to their charge-shift character. The bonding between homonuclear atoms in the studied molecules was described as a resonance between three VB structures (Figure~\ref{fig:VBstructures}). The accuracy of the employed VB and VB-QMC methods was first tested in calculations of bond dissociation energies. The obtained bond dissociation energies were compared with the corresponding CCSD(T) and experimental values (Table~\ref{tab:dissociation}). 
There is reasonable good agreement between the VB-VMC and SD-BOVB results. According to the mean relative error in the studied series, the VB-VMC method gives overall even slightly better results than the CCSD(T) method. The VB-DMC method provides the most accurate calculation of the bond dissociation energies and gives the best agreement with the experimental results. 

\begin{table}
\caption{Bond dissociation energies (in kcal/mol) and mean relative errors (MRE, in ${\%}$) calculated by different methods with respect to the corresponding experimental values.}
\label{tab:dissociation}
\begin{tabular}{ |c|c|c|c|c|c|c| } 
 \hline
                       & SD-BOVB & VBCISD & VB-VMC & VB-DMC & CCSD(T) & Exp.$^a$ \\
 \hline
  F${_2}$              & 34.7    & 33.8   & 26.4   & 34.1   & 30.4    & 38.3  \\
  Cl${_2}$             & 47.3    & 51.7   & 55.0   & 57.3   & 45.8    & 58.0  \\
  Br${_2}$             & 39.2    & 43.9   & 43.6   & 50.6   & 39.3    & 45.9  \\
  H${_3}$C-CH${_3}$    &  91.4   &        & 92.0   & 96.5   & 93.0    & 96.7  \\
  H${_2}$N-NH${_2}$    &  66.2   &        & 65.2   & 69.6   & 64.8    & 75.4  \\
  HO-OH	               &  52.0   &        & 41.1   & 49.4   & 47.8    & 53.9  \\
  \hline
  MRE	               &  10.61\%   & 8.99\%   & 13.90\%  & 6.44\%   & 14.21\%   &       \\
 \hline
\multicolumn{7}{l}{$^a$Experimental values from Ref.~\onlinecite{NIST-BOOK-15}.}
\end{tabular}
\end{table}

Table~\ref{CCweights} displays the weights of the VB structures presented in Figure~\ref{fig:VBstructures} obtained with the SD-BOVB, VBCISD, and VB-VMC methods. It can be seen that these approaches give quite similar weights. For most of the studied molecules the calculated VB-VMC weights of the ionic structures are somewhat greater compared to the corresponding SD-BOVB and VBCISD weights. This trend is expected because in ionic structures there are more electrons concentrated on specific atomic regions (i.e. the atoms bearing a negative charge), thus leading to a larger correlation effect in ionic structures as compared with covalent structures, which cannot be fully accounted for by the SD-BOVB method (and even by truncated CI) where only the differential dynamical correlation necessary to provide accurate energy differences is included. By contrast, the Jastrow factor used in VB-VMC, which describe the correlation effect through an explicit treatment in an equal manner for all electrons, is logically able to retrieve the larger correlation of the ionic structures, thus leading to larger ionic weights as compared with the traditional correlated VB methods.

\begin{table}
\caption{Chirgwin-Coulson weights (in ${\%}$) of the VB structures 1/2/3 shown in Figure~\ref{fig:VBstructures} for the series of studied molecules.}
\label{CCweights}
\begin{tabular}{ |c|c|c|c| } 
 \hline
                     & SD-BOVB        & VBCISD         & VB-VMC \\
 \hline
  F${_2}$            & 67.3/16.4/16.4 & 71.2/14.4/14.4 & 68.2/15.9/15.9 \\
  Cl${_2}$           & 64.0/18.0/18.0 & 66.0/17.0/17.0 & 59.8/20.1/20.1 \\
  Br${_2}$           & 68.8/15.6/15.6 & 68.4/15.8/15.8 & 60.6/19.7/19.7 \\
  H${_3}$C-CH${_3}$  &  51.2/24.4/24.4 &                & 55.4/22.3/22.3 \\
  H${_2}$N-NH${_2}$  &  55.6/22.2/22.2 &                & 57.6/21.2/21.2 \\
  HO-OH              &  63.0/18.5/18.5 &                & 62.8/18.6/18.6 \\
 \hline
\end{tabular}
\end{table}

The calculated charge-shift resonance energies (RE$_{\text{CS}}$) for the examined series are collected in Table~\ref{tab:resonance}. It should be noted that the calculated SD-BOVB RE$_{\text{CS}}$ values are very close to the ones published by Shaik {\it et al.}~\cite{Shaik-CSB-NaturChem-2009, Shaik-CBS-Chem_EurJ-2005}. The charge-shift resonance energies calculated by means of the VB-VMC method are very close to the SD-BOVB ones. In the series of the dihalogens the calculated VB-VMC RE$_{\text{CS}}$ are in between the values obtained by SD-BOVB and VBCISD. The VB-VMC method improves the results of the SD-BOVB approach, since it is known that the latter overestimates the contribution of ionic structures to the resonance, resulting in somewhat smaller RE$_{\text{CS}}$ values. The VB-DMC approach provides markedly lower RE$_{\text{CS}}$ values for the studied bonds than the reference VBCISD and SD-BOVB methods. This is expected as the DMC algorithm operates a projection of the given trial wave function onto the ground-state FN wave function, and therefore is not suited to calculate diabatic states, such as a separate covalent or ionic structure, which will at least partly lose their specific identity at the VB-DMC level. In principle, we could have the same problem in VB-VMC since VMC becomes theoretically equivalent to DMC for a fully flexible Jastrow factor. However, in practice, the form of the Jastrow factor that we use is not flexible enough to alter the nature of diabatic states. Therefore, the present VB-VMC method can be used to compute diabatic states, and extract quantities related to it such as resonance energies.

\begin{table}
\caption{Charge-shift resonance energies (in kcal/mol).}
\label{tab:resonance}
\begin{tabular}{ |c|c|c|c|c| } 
 \hline
                       & SD-BOVB & VBCISD & VB-VMC & VB-DMC \\
 \hline
  F${_2}$              & 77.4    & 70.7   & 73.4   & 52.5 \\
  Cl${_2}$             & 44.9    & 34.3   & 42.1   & 21.9 \\
  Br${_2}$             & 39.4    & 28.6   & 33.7   & 16.1 \\
  H${_3}$C-CH${_3}$    &  23.5    &        & 22.6   & 8.1 \\
  H${_2}$N-NH${_2}$    &  42.1    &        & 43.6   & 19.6 \\
  HO-OH	               &  68.5    &        & 65.5   & 42.1 \\
 \hline
\end{tabular}
\end{table}

\subsubsection{Resonance model for allyl systems}

Allyl systems represent the smallest ${\pi}$-resonant molecules. Despite their simple structure, the resonance model of the allyl systems has been a subject of interest of theoretical chemists in the past decades. The evaluation of the resonance energies in allyl conjugated systems is another challenging task that raised an ongoing debate.~\cite{G0bbi-Allyl-JACS-1994, Mo-Allyl-JPC-1996, Mo-delocal-JCP-1998, Wiberg-Allyl-JACS-1990} In previous papers~\cite{Linares-Allyl-JPCA-2006, Linares-Allyl-JPCA-2008} published by one of the present authors, the resonance energies of allyl systems were calculated at the levels of modern VB theory, and, in addition, a theoretical model was proposed to describe the nature of resonance in these systems. 

\begin{figure}[htp]
    \centering
    \includegraphics[width=15cm]{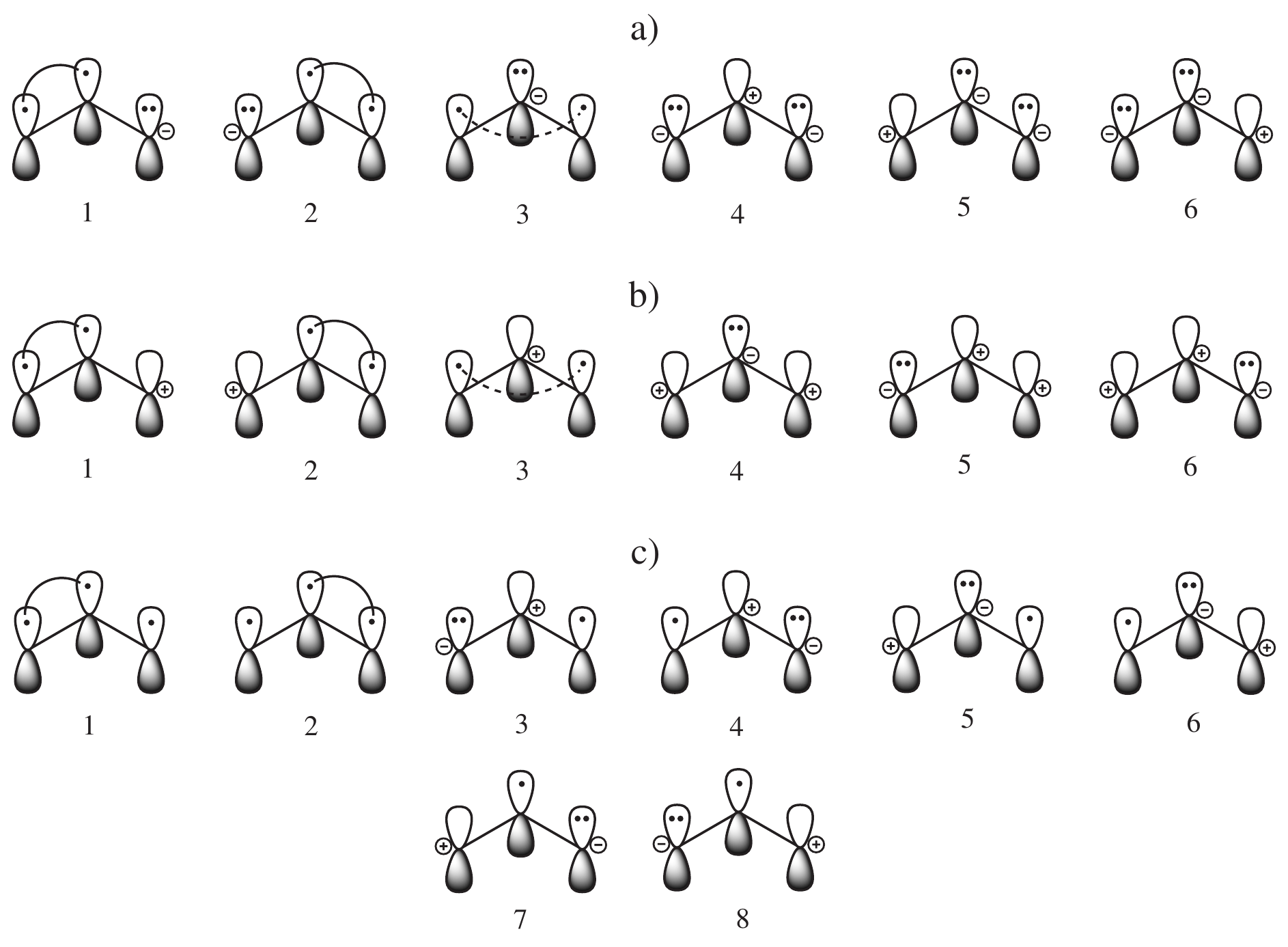}
    \caption{Set of VB structures for the different allyl systems: a) anion, b) cation, and c) radical.}
    \label{fig:VBstructures_allyl}
\end{figure}

In the present work, the ${\pi}$-electron system of the studied molecules was described through the set of six and eight VB structures for the allyl anion, cation, and radical, respectively (Figure~\ref{fig:VBstructures_allyl}). Note that, in Ref.~\onlinecite{Linares-Allyl-JPCA-2008}, the allyl ions were represented by three VB structures (structures 1, 2, and 3 in Figure~\ref{fig:VBstructures_allyl} a and b), whereas the allyl radical was described by two VB structures (structures 1 and 2 in Figure~\ref{fig:VBstructures_allyl} c) because the structures were built using Coulson-Fischer-type active orbitals, leading to a description where classical ionic VB structures are implicitly included, and where one VB structure maps to one specific bonding pattern (or ``Lewis structure''). Table~\ref{tab:CCweights_allyl} provides the weights of the VB structures presented in Figure~\ref{fig:VBstructures_allyl} obtained by means of the BOVB (with delocalized active orbitals) and VB-VMC methods. Again, these two approaches give very similar weights. It should be noted that some of the VB structures that were not explicitly considered in Ref.~\onlinecite{Linares-Allyl-JPCA-2008} can have considerable weights within a classical VB description. In the case of the allyl ions, these are the symmetrical structures 4 (Figure~\ref{fig:VBstructures_allyl} a and b). For the allyl radical, these are the two ionic structures 3 and 4 which have weights over 10${\%}$.

\begin{table}
\caption{Chirgwin-Coulson weights (in {\%}) of the VB structures in Figure~\ref{fig:VBstructures_allyl} for the allyl systems.}
\label{tab:CCweights_allyl}
\begin{tabular}{ |c|c|c|c|c|c|c| } 
   \hline
  & \multicolumn{2}{c|}{anion} & \multicolumn{2}{c|}{cation} & \multicolumn{2}{c|}{radical} \\
  \hline
 Structure & BOVB & VB-VMC & BOVB & VB-VMC & BOVB & VB-VMC \\                      
  \hline
1 & \ 29.30 & 28.81 & \ 28.36 & 27.98 & \ 29.81 & 28.35 \\
2 & \ 29.30 & 28.81 & \ 28.36 & 27.98 & \ 29.81 & 28.35 \\
3 & \ 11.71 & 10.39 & \ 20.71 & 22.15 & \ 10.47 & 12.08 \\
4 & \ 26.44 & 27.69 & \ 18.92 & 15.82 & \ 10.47 & 12.08 \\
5 & \ 1.62  & 2.14  & \ 1.83  & 3.04  & \ 7.42  & 6.70  \\
6 & \ 1.62  & 2.14  & \ 1.83  & 3.04  & \ 7.42  & 6.70  \\
7 &         &       &         &	      & \ 2.29  & 2.87 \\
8 &         &       &         &       & \ 2.29  & 2.87 \\
 \hline
\end{tabular}
\end{table}

The calculated resonance energies of the studied molecules are reported in Table~\ref{tab:resonance_allyl}. The resonance energy was calculated relative to the system described by three VB structures. For instance, for the allyl cation, the reference structure was described as a resonance between the VB structures 1, 4, and 5 (Figure~\ref{fig:VBstructures_allyl} b). This way of calculating the resonance energy is based on the idea that each covalent bond in a Lewis structure can be expanded in its ionic and covalent components. As can be seen from the data given in Table~\ref{tab:resonance_allyl}, the values obtained by means of BOVB method using the extended set of VB structures are very similar as the ones given in Ref.~\onlinecite{Linares-Allyl-JPCA-2008}. The VB-VMC resonance energies are for all allyl systems somewhat smaller than the BOVB ones. This is because, as already mentioned, BOVB may underestimate resonance energies to some extend, because the full-structure BOVB wave function includes static and some dynamical correlation while the single-structure computed separately corresponding to the  ``non-resonant'' reference situation is totally devoid of any correlation and therefore its energy is overestimated as compared with the full-structure BOVB wave function. By constrast, the VB-VMC method, through the explicit correlated treatment ensured by the Jastrow factor, enables the inclusion of electron correlation in the full-structure (adiabatic) wave function and in the separate (``non-resonant'' reference) structure alike, therefore leading to lower and in principle more accurate resonance energies. In terms of chemical trends, it is found that resonance in allyl ions is particularly strong, with the cation being more resonant than the anion, whereas the allyl radical is significantly less resonant compared to the corresponding ions.

\begin{table}
\caption{Resonance energies (in kcal/mol) for the allyl systems.}
\label{tab:resonance_allyl}
\begin{tabular}{ |c|c|c| } 
 \hline
          & BOVB & VB-VMC \\
 \hline
 anion    & 48.2   & 42.6 \\
 cation   & 54.2   & 48.4 \\
 radical  & 31.2   & 29.7 \\
 \hline
\end{tabular}
\end{table}

\subsubsection{Resonance model for some carbonyl compounds}

Chemical properties of carboxylic acids and their derivatives is one of the most fundamental topics in organic chemistry. It is well known that different carbonyl-based functional groups can have very different chemical properties. In particular, formyl chlorides are known to be very reactive under a nucleophilic attack, whereas amides are much less and esters lie in between. Here, we consider a series of protonated carbonyl compounds: formyl chloride ({\bf 1}), methyl formate ({\bf 2}), and formamide ({\bf 3}). The reactivity of the studied carbonyl molecules under a nucleophilic attack is given as ${\bf {1} > \bf {2} > \bf {3}}$. Such order of reactivity is traditionally explained through a stabilizing resonance interaction in the reactant molecule that goes under the nucleophilic attack.

\begin{figure}[t]
    \centering
    \includegraphics[width=15cm]{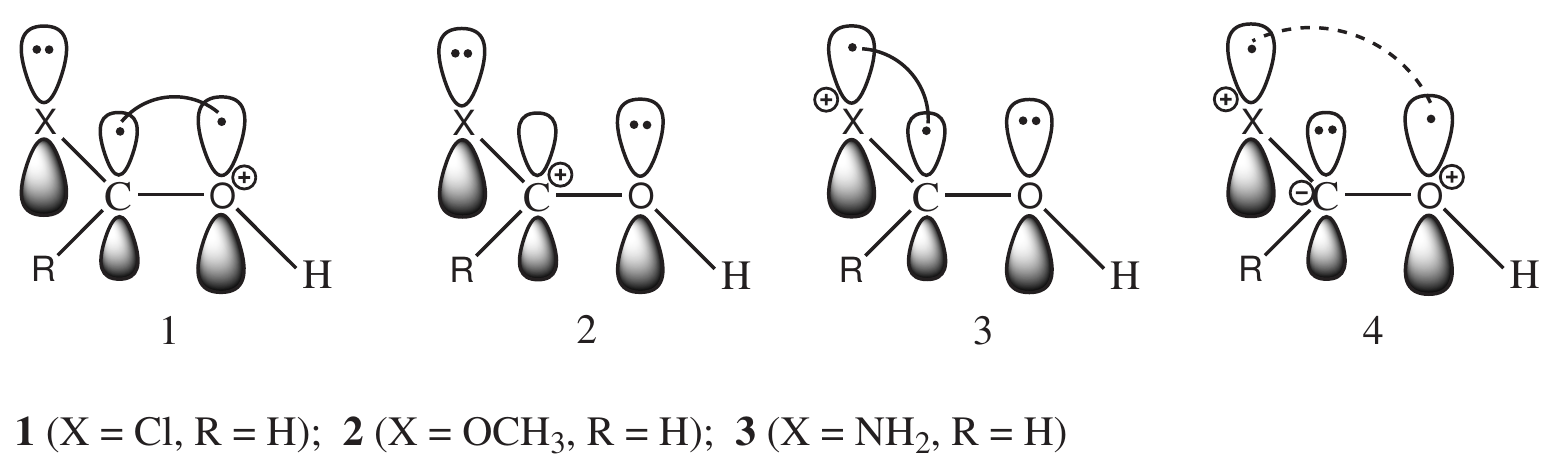}
    \caption{VB structures for protonated formyl chloride ({\bf 1}), methyl formate ({\bf 2}), and formamide ({\bf 3}).}
    \label{fig:VBstructures_form}
\end{figure}

The resonance in the studied molecules was described through the set of four VB structures as depicted in Figure~\ref{fig:VBstructures_form}. The weights of the VB structures for the studied molecules are reported in Table~\ref{tab:CCweights_form}. The main difference between the results of BOVB (with delocalized active orbitals) and VB-VMC are found for the weights of the VB structure 2. According to the VB-VMC results, structure 2 is the most relevant one in the description of the ${\pi}$-electron system of the examined molecules. This point will be examined and commented in further details in a forthcoming article.

\begin{table}
\caption{Chirgwin-Coulson weights (in {\%}) of the VB structures in Figure~\ref{fig:VBstructures_form} for protonated formyl chloride ({\bf 1}), methyl formate ({\bf 2}), and formamide ({\bf 3}).}
\label{tab:CCweights_form}
\begin{tabular}{ |c|c|c|c|c|c|c| } 
 \hline
  & \multicolumn{2}{c|}{molecule {\bf 1}} & \multicolumn{2}{c|}{molecule {\bf 2}} & \multicolumn{2}{c|}{molecule {\bf 3}} \\
  \hline
 Structure & BOVB & VB-VMC & BOVB & VB-VMC & BOVB & VB-VMC \\                      
  \hline
1 & \ 36.6 & 28.6 & \ 25.6 & 18.8 & \ 22.4 & 16.4 \\
2 & \ 37.6 & 51.7 & \ 34.0 & 49.9 & \ 32.0 & 46.5 \\
3 & \ 20.9 & 16.9 & \ 27.7 & 27.7 & \ 39.5 & 33.3 \\
4 & \  4.9 &  2.8 & \  5.6 &  3.6 & \  6.2 &  3.8 \\
 \hline
\end{tabular}
\end{table}

\begin{figure}[t]
    \centering
    \includegraphics[width=15cm]{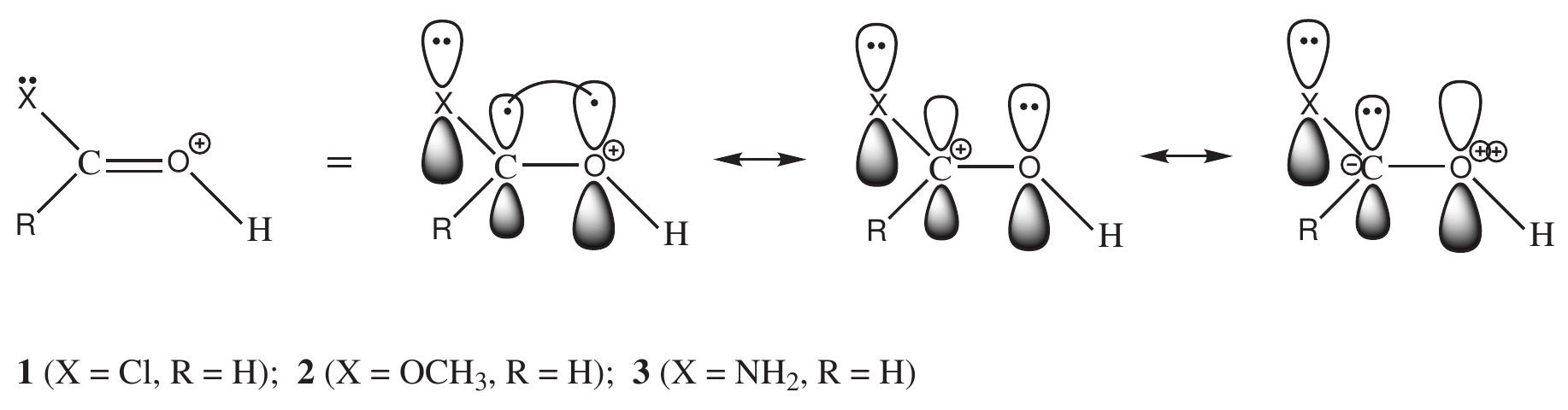}
    \caption{Expansion of one Lewis structure (structure 1 in Figure~\ref{fig:VBstructures_form}) in terms of the three Pauling covalent-ionic structures.}
    \label{fig:VBstructures_carbonyl}
\end{figure}

The resonance energies for the examined protonated carbonyl compounds were calculated relative to the system described by three VB structures (Figure~\ref{fig:VBstructures_carbonyl}). The values of resonance energies given in Table~\ref{tab:resonance_carbonyl} show that the stabilization of the studied molecules due to resonance increases in the series of molecules ${\bf {1} - \bf {3}}$.  This is in agreement with the traditional explanation of the reactivity order in the studied series.

All considered, it has been found in this section that the VB-VMC method retrieves all the interpretative capabilities of the more traditional BOVB method, together with at least a similar or better accuracy as compared with the highest level SD-BOVB and VBCISD reference methods. In particular, the VB-VMC method allows the calculation of structure weights and resonance energies, with an improved description of ionic structures and by computing separate structures associated with a given reference diabatic state with the full inclusion of electron correlation. The VB-DMC method leads on its side to the highest accuracy for energy differences, however it is not suited for computing diabatic states in general, as it cannot preserve the chemical identity of a given structure (see discussion at the end of Section~\ref{sec:csreso}).

\vspace{5mm}

\begin{table}
\caption{Resonance energies (in kcal/mol) for protonated formyl chloride ({\bf 1}), methyl formate ({\bf 2}), and formamide ({\bf 3}).}
\label{tab:resonance_carbonyl}
\begin{tabular}{ |c|c|c| } 
 \hline
          & BOVB & VB-VMC \\
 \hline
 molecule {\bf 1}    & 24.6   & 24.2 \\
 molecule {\bf 2}    & 56.3   & 46.5 \\
 molecule {\bf 3}    & 60.8   & 52.0 \\
 \hline
\end{tabular}
\end{table}

\section{Applications of the Valence-Bond Quantum Monte Carlo method}
\label{sec:applications}

\subsection{The V state of ethylene}

The accurate calculation of the vertical excitation energy (VEE) of ethylene from its so-called ``N'' ground to its ``V'' excited state has been for decades a particularly challenging issue. Many theoretical attempts have been made that span no less than four decades, from the early studies of Goddard \textit{et al.} to the more recent work of Angeli (see Ref.~\onlinecite{WuZhaBraShaHib-TCA-14} and references therein). CI calculations require an incredibly large number of determinants (many billions) for such a small molecule to obtain a converged VEE, while, as far as multi-configurational methods are concerned, even multireference second-order perturbation theory (CASPT2) with an active space made of 12 electrons in 12 orbitals has proved to be unsuccessful~\cite{Ang-JCC-09}.

\begin{figure}[t]
    \centering
    \includegraphics[width=10cm]{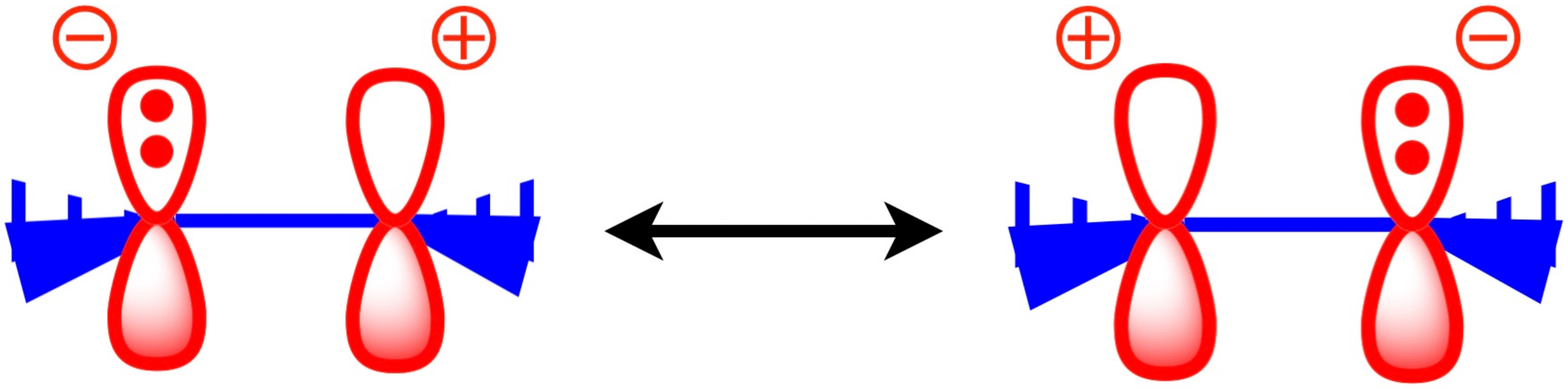}
    \caption{The conventional Lewis description of the “ionic” V state of ethylene.}
    \label{fig:scheme1ethylene}
\end{figure}

The issue is that, to obtain an accurate VEE, one has to describe both the ground N state and excited V state in a balanced way, as far as the inclusion of electron correlation is concerned. The ground state displays a standard double C-C bond and sigma C–H bonds. The V state, however, belongs to the family of dipolar $\pi$ ``ionic states'', which could be described in a conventional way by the resonance of two zwitterionic structures, as depicted in Figure~\ref{fig:scheme1ethylene} (as we will see in the following, the precise physical nature of this V state is actually more complex). Therefore, this state is characterized by a particularly strong electronic fluctuation, and it is stabilized by the large resonance arising between the two structures. To properly describe this state, several issues have to be considered. First, it is important to properly describe the dynamical correlation of the active $\pi$ electron pair, which is particularly important in the V state where the pair is concentrated alternatively on one of the two carbon atoms, while the two $\pi$ electrons are spatially more separated in the covalent bonded ground state. A second important issue for the excited V state wave function is to properly account for the dynamic response of the $\sigma$ skeleton to the fluctuation of the $\pi$ electrons. When the active $\pi$ electrons fluctuate from one ionic situation to the other, the $\sigma$ C–C and C–H bonds shall follow this fluctuation by repolarizing themselves accordingly. This effect, which is sometimes called ``dynamic $\sigma$ repolarization'', is particularly difficult to include in post-HF treatments, as it appears for instance at the fourth order of perturbation theory only.~\cite{Ang-JCC-09} This explains why a full-valence CASPT2 fails to properly describe this state. Another issue is that, at both HF and CASSCF levels, the V state comes out close in energy to a Rydberg intruder state, leading to a spurious mixing of the two states which produces optimized orbitals that are way too diffuse for the V state.~\cite{WuZhaBraShaHib-TCA-14} Overall, these particularities make a balanced description of the two N and V states very difficult to achieve using standard post-HF and post-CASSCF correlation treatments, and in such a framework insanely large CI expansions or high levels of perturbation treatments are necessary in a brute-force approach. As will be seen below, VB theory offers an elegant and efficient framework to tackle this challenge, using very compact wave functions.

\begin{figure}[htp]
    \centering
    \includegraphics[width=10cm]{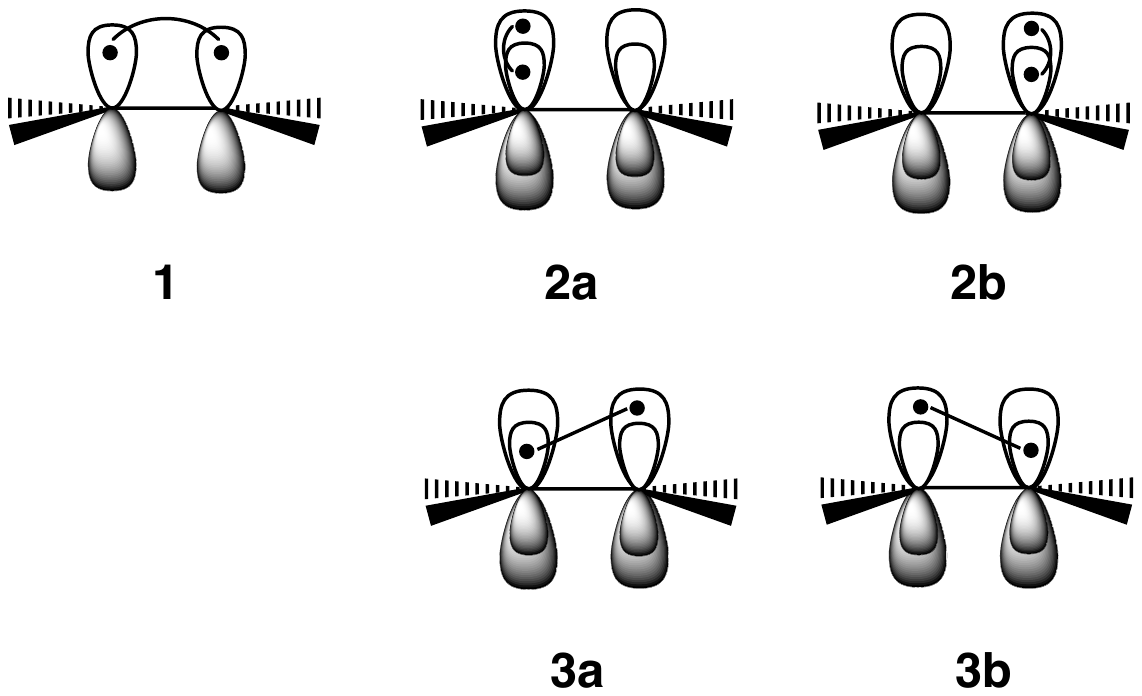}
    \caption{The basis of VB structures for the description of the N and V states of ethylene.}
    \label{fig:scheme2ethylene}
\end{figure}

Let us derive the basis of VB structures needed to properly describe the N $\to$ V excitation. The main difference between the two states comes from the configuration of the two $\pi$ electrons, while the $\sigma$ frame remains basically unchanged. Therefore, only the two $\pi$ electrons shall in principle be considered as active if one is interested in an accurate estimation of the VEE. As previously explained, an accurate treatment of the correlation of the $\pi$ electron pairs in the ionic structure is required to reach a proper description of the V state. This requirement can be fulfilled in the classical VB framework by ``splitting'' the $\pi$ orbitals into two pairs of active orbitals, with each pair localized on a specific carbon atom, one orbital of the pair being more diffuse and the second one more contracted. Then, in each ionic structure, the two active $\pi$ electrons are therefore described as a pair of spin-singlet-coupled electrons in two (localized) orbitals, similarly as in the covalent structure, except that in the case of the ionic structures the two active orbitals are localized on the same carbon atom. All in all, the situation corresponds to two active electrons in four active orbitals, leading to the complete and non-redundant set of five structures associated with an overall singlet state that are displayed in Figure~\ref{fig:scheme2ethylene}. The first three structures (1, 2a, 2b) are the traditional covalent and ionic structures that mix into the N ground state. Structures 3a and 3b may appear as quite surprising at first glance, however these are legitimate structures in the ``split'' classical VB framework, and correspond to ``asymmetric'' covalent structures. A combination of these two structures of adequate symmetry mixes into the ionic V state, which shall then be described by a resonance of four structures only (2a, 2b, 3a, 3b).

\begin{table}
\caption{N $\to$ V vertical excitation energy of ethylene (in eV), at different levels of theory using an ``aug-VTZ'' basis set, together with the estimated exact value, all taken from Ref.~\onlinecite{WuZhaBraShaHib-TCA-14} (see this reference for more details). The number used as suffix (3, 5 or 15) corresponds to the number of structures in the complete basis of VB structures used to describe both the N and V states of ethylene. }
\label{tab:ethylene}
\begin{tabular}{|c|c|} 
\hline
     & Excitation energy   \\
 \hline
 CASSCF & 8.48 \\
 BOVB-5 & 8.01 \\
 BOVB-15 & 7.97 \\
 VB-VMC-5 & 8.59(2)\\
 BOVB-VMC-5 & 7.96(2)\\
 BOVB-VMC-3 & 8.98(2)\\
  BOVB-DMC-5 & 7.93(1)\\
  Estimated exact & 7.88\\
 \hline
\end{tabular}
\end{table}

As explained before, a physically correct description of the V electronic state requires both dynamical correlation of the active electron pair and the so-called dynamic $\sigma$ repolarization effect to be included. An efficient way to do so, which also avoids the spurious Rydberg mixing leading to too diffuse active orbitals after optimization, is to use a level of theory able to include these two effects right at the orbital optimization level. This could be achieved either by using the BOVB method (in its highest SD-BOVB variant), or by using VB-QMC. The SD-BOVB method indeed makes use of different orbitals for different structures, enabling the dynamic $\sigma$ repolarization effect to be included at the orbital level, by having the $\sigma$ system that polarizes in an opposite way in the two ionic structures 2a and 2b as a result of the orbital optimization process. 

Estimates of the N $\to$ V vertical excitation energies at different levels of theory are reported in Table~\ref{tab:ethylene}. When using the five active orbitals displayed in Figure~\ref{fig:scheme1ethylene}, the (SD-)BOVB method achieves a VEE of 8.01 eV, in good agreement with the best theoretical estimate of 7.88 eV, which is a particularly impressive achievement for such compact wave functions (only 3 structures of the N state and 4 structures for the V state). A slightly improved estimate of 7.97 eV could even be obtained when the valence $\sigma$ electrons are also considered as active and treated at the VB level (leading to a complete set of 15 VB structures). Let us now consider the different VB-QMC results. In all wave functions, both the Jastrow parameters, CI expansion coefficients, and orbitals are reoptimized in VMC. The VB-VMC method using the 5 structures displayed in Figure~\ref{fig:scheme2ethylene} produces a poor estimate of 8.59 eV, showing that the Jastrow function used cannot properly account for the strong dynamic $\sigma$ repolarization effect in the V state. When a BOVB determinantal part is used and supplemented by a Jastrow factor (BOVB-VMC-5 level), the VEE is dramatically improved to 7.96 eV, comparable within the statistical error bars to the best SD-BOVB estimate (``BOVB-15'' in Table~\ref{tab:ethylene}). Quite interesting is the fact that the energy of the V state rises by about 1 eV when structures 3a and 3b are not included in the BOVB-VMC wave function (BOVB-VMC-3 vs. BOVB-VMC-5 values in Table~\ref{tab:ethylene}), illustrating the critical importance of these two asymmetric covalent structures to reach a quantitative description of the V state. Their importance also shows up in their significantly large weights, almost overall 20\% in the Löwdin definition. This teaches us that the V state of ethylene is not a pure ionic state as usually believed, but also includes a secondary but significant covalent component. Last, when a FN ground-state projection is performed by the DMC algorithm using the very compact BOVB-VMC-5 wave function as trial wave function, an improved estimate of 7.93 eV is obtained (BOVB-DMC-5 entry in Table~\ref{tab:ethylene}), very close to the best theoretical estimate of 7.88 eV. 

\subsection{Multicenter bonding in the ditetracyanoethylene dianion and ``pancake bonding'' systems}

VB-QMC methods were used to investigate the nature of the bonding in the tetracyanoethylenyl anion radical dimer, ((CN)$_4$C$_2$)$_2^{2-}$. \cite{10.1021/ct400290n} Tetracyanoethylene (TCNE) is a widely used model for organic superconductors and is a well-known strong $\pi$-acid that reacts at room temperature with many metals including aluminum, copper, magnesium, sodium, and potassium.\cite{doi:10.1021/ja00878a017} The resulting salts result in parallel stacked tetracyanoethylenyl anions with the metal cations in equatorial positions (in between the layers). The crystalline solid states can be modeled using the stacked dianionic dimer which exhibits interesting long-range multi-center carbon-carbon bonding and can be used as a prototypical system for understanding a particular large family of stacked radical dimers, also called ``pancake bonding'' systems. While the described configuration would suggest that electrostatics are the primary force holding the dimers together, the observation that changing the size of the cation, that is in between the two layers, does not change the interlayer distance of 
2.90$\pm$0.05 {\AA} strongly suggests that other forces hold the dimer together as well.

A computational study by Jung and Head-Gordon\cite{B403450C} provided proof that long-range carbon-carbon bonding interactions were present in the TCNE dimer. Rather than studying the more stable configuration with equatorial cations, they studied the dimer with cations on opposite sides of the dimer pancake (axial positions). This configuration eliminated the attractive electrostatics and left only repulsive electrostatics between the dimers. If electrostatics were the only force stabilizing the dimer, the axial configurations would be unstable. They found the binding energy of the axial configuration to be 11.2 kcal/mol with a counter-poise corrected multi-configuration quasi-degenerate perturbation theory (MCQDPT) calculation on top of a CASSCF wave function with an active space made of 2 electrons in 2 orbitals. Interestingly, without the perturbative corrections the CASSCF level found the axial configurations to be unstable, which suggests that the bonding interaction is not a simple carbon-carbon covalent bond. However, they found a Bader bond (3,-1) critical point in the region between the carbons of the separate TCNE monomers, which would suggest, under this analysis, that some type of chemical bonding takes place between the carbons of the dimer. They, along with Garc\'ia-Yoldi \textit{et al.}~\cite{ https://doi.org/10.1002/jcc.20525}, interpreted the importance of dynamical correlation in stabilizing the dimers in terms of an attractive dispersion interaction between the faces of the dimer and some $\pi$-$\pi$ inter-dimer interactions. However, such a strong dispersion interaction that would overbalance a strong repulsive electrostatic interaction in the DTCNE dimer with cations on the opposite side would be unprecedented on such a short $\pi$-staking system. Besides, the short inter-dimer carbon-carbon distance of 2.9 {\AA} is not compatible with a pure dispersion interaction, as the sum of van der Waals radii corresponds to a significantly large distance of 3.4 {\AA}. The lack of insight into the exact nature of the covalent bonding between the dimers stimulated a re-examination of the system with VB-QMC techniques.

\begin{figure}[t]
    \centering
    \includegraphics[width=15cm]{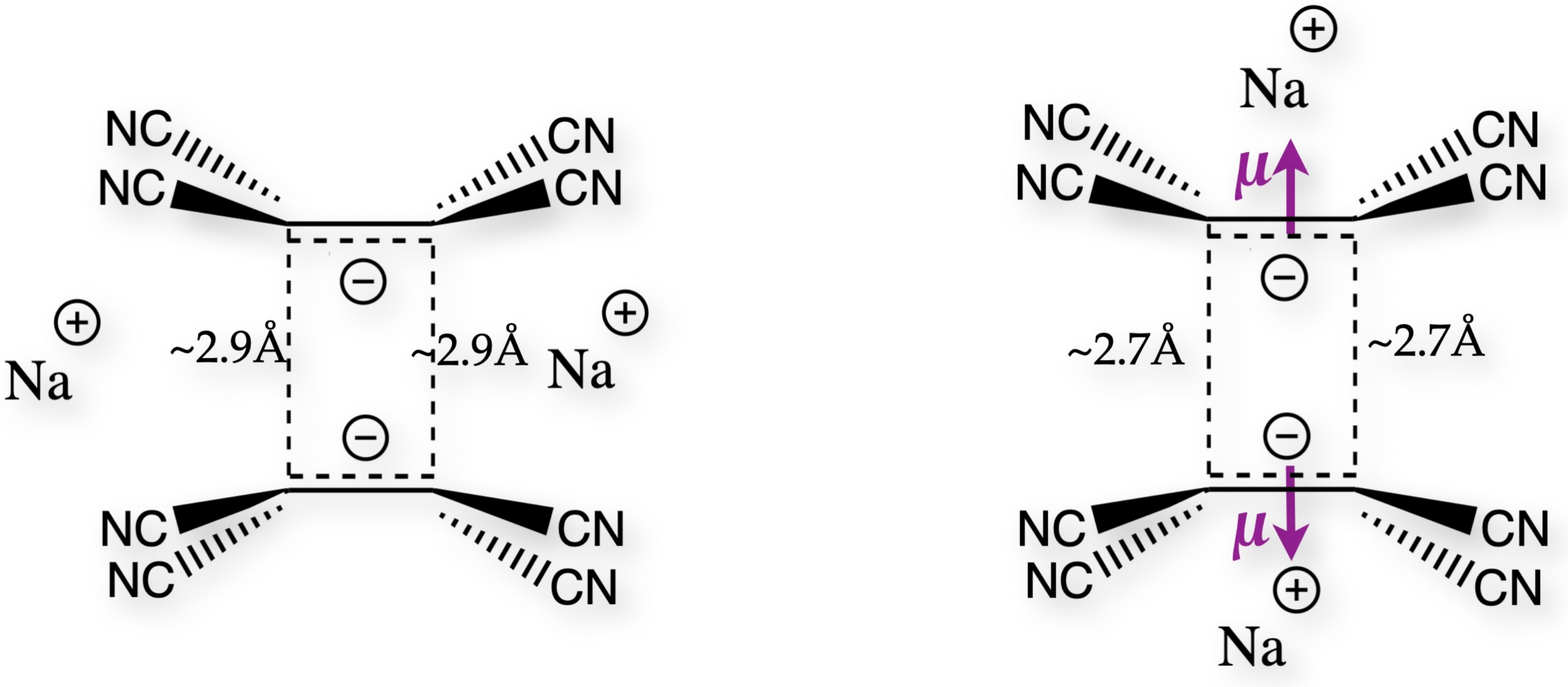}
    \caption{Schematic representation of the most stable configuration of the (TCNE$^-$;Na$^+$)$_2$ system with cation in equatorial position (left) and of the studied configuration with cations in axial position leading to unfavorable electrostatic interactions (right).}
    \label{fig:TCNE_scheme1}
\end{figure}

Both axial and equatorial configurations of sodium tetracyanoethylene dimers were investigated using CCSD(T) and VB methods. The geometries were optimized with MP2 calculations using a double-zeta basis set followed by a re-optimization of the inter-fragment distance using counter-poise corrected CCSD(T) calculations.  A minimum geometry was found at an inter-fragment distance of 2.56 {\AA}, and its dissociation to infinite separation of the fragments was found to require an energy of 11.6 kcal/mol. The T1-diagnostic for the CCSD(T) calculations was well below the critical value of 0.02 (a T1-diagnostic above this critical value may indicate unreliable CCSD(T) calculations and a strong multi-configurational character). Both geometries and binding energies were reasonably consistent with the calculations of Jung and Head-Gordon~\cite{B403450C}. For the axial configurations (Figure~\ref{fig:TCNE_scheme1}, right) the cyano groups slightly pucker away from the cyano groups of the monomer, something that is also seen in X-ray structures of crystalline TCNE and which shows up is the fact that the overlap between cyano fragment orbitals across the carbon-carbon double bond within a monomer is comparable to the overlap between cyano fragment orbitals of the two separate monomers. 

\begin{figure}[t]
    \centering
    \includegraphics[width=15cm]{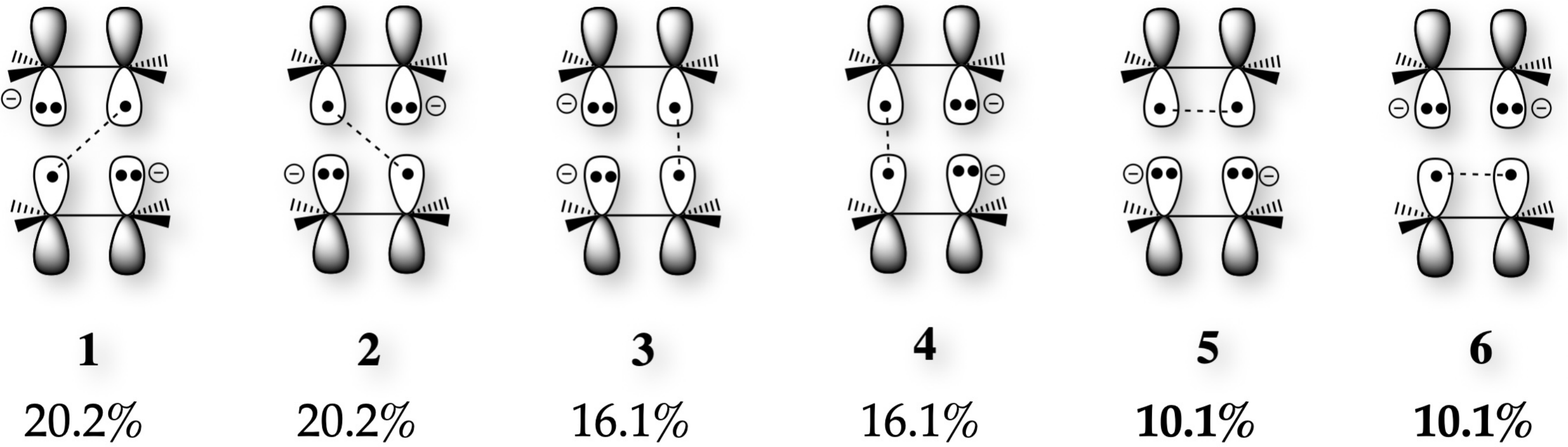}
    \caption{The main six VB structures describing bonding in the DTCNE dimer, together with the computed VB-VMC Chirgwin-Coulson weights with a 10-structure wave function.}
    \label{fig:TCNE_scheme2}
\end{figure}

All VB-QMC calculations and analysis are for the geometry on the right in Figure~\ref{fig:TCNE_scheme1}, because, since the electrostatic interaction is repulsive in this configuration, this allows one to pinpoint the very nature of the chemical component of the bonding taking place between the two monomers. The ground-state VB-VMC wave function was calculated using the complete set of ten VB structures that arises from the distribution of six active electrons in four active orbitals with an overall spin-singlet coupling, the active orbitals being the four central carbons p-type orbitals that point towards each other (and therefore can overlap significantly) and thus lead to chemical bonding between the two monomers. Figure~\ref{fig:TCNE_scheme2} displays the main six structures together with their computed VB-VMC weights (the weights of the remaining four structures, which are very unfavorable multi-ionic structures, sum up to about 7\%). These six VB structures overall describe an aromatic four-center six-electron type interaction, quite akin to the aromatic S$_2$N$_2$ and related S$_2$E$_2$ and E$_4^{2+}$ systems (E = S, Se, Te)~\cite{BraLoHib-CPC-12}. The VB structures were simultaneously optimized with the Jastrow factor in order to provide dynamical correlation that is completely absent in pure VBSCF wave functions.  All six main structures display important weights in the 10-20\% range, and therefore describe a situation of large electron fluctuation accompanied with a shift of formal charges from one VB configuration to another, quite reminiscent of the ``charge-shift'' bonding type of interaction. It is important to note that structures 1-4 describe a fluctuating situation with three active electrons on each monomer, which, considering their almost equal weights, correspond to the description of two $\pi$-type 2-center 3-electron bonds between the two central carbon atoms in each monomer. A two-center three-electron bond is a type of bond that was described for the first time by Pauling in 1931~\cite{Pau-JACS-31,Pau-JCP-33}, and which has been shown lately to constitute, with the two-center one-electron bond, a particular category within the ``charge-shift'' family of bonding~\cite{ShaDanGalBraWuHib-ACIE-20,ShaDanGalBraWuHib-AC-20}. 

\begin{figure}[t]
    \centering
    \includegraphics[width=15cm]{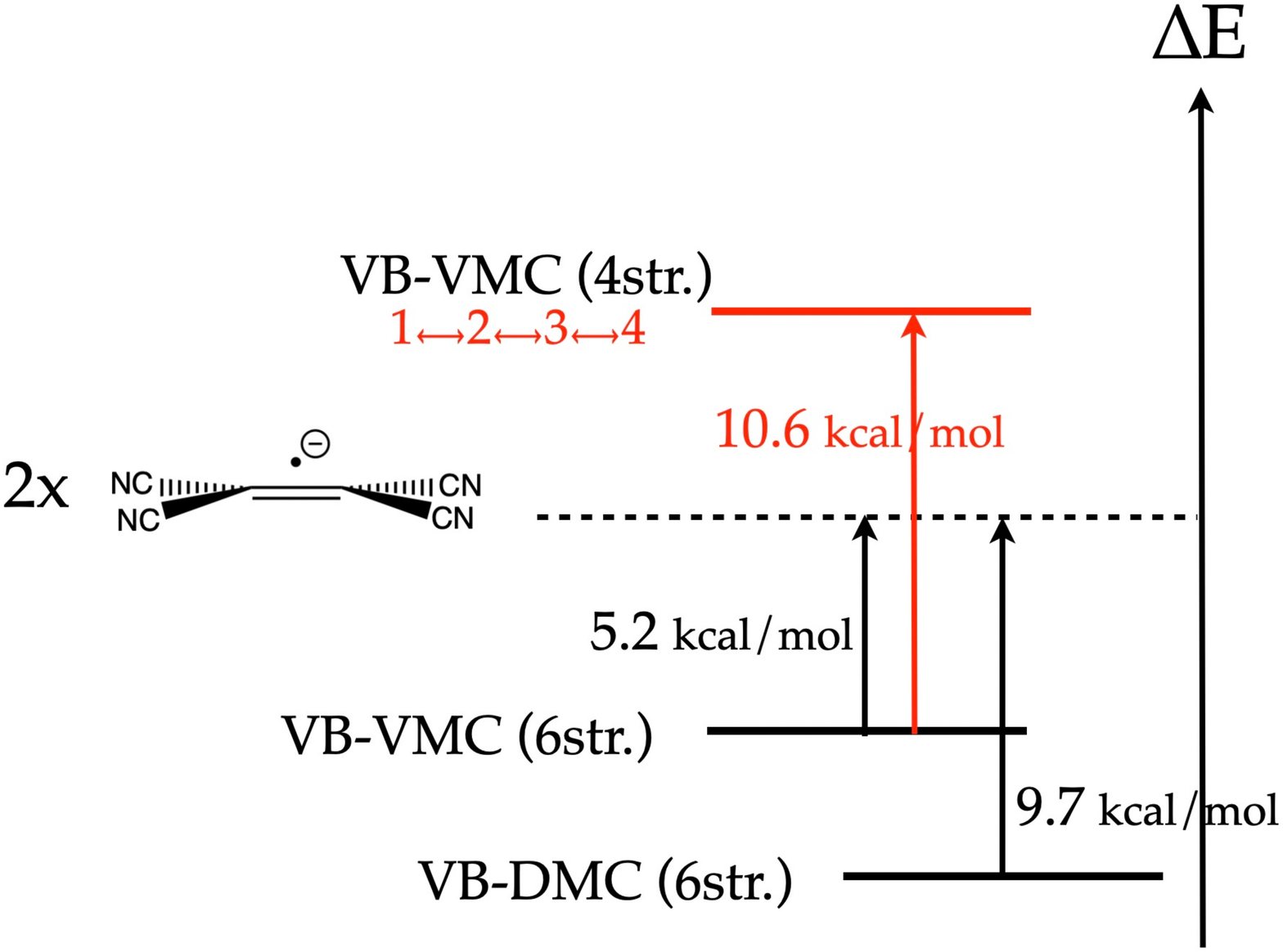}
    \caption{Energy differences between the different VB-QMC calculations for the DTCNE dimer.}
    \label{fig:TCNE_scheme3}
\end{figure}

At the VB-VMC level, when a separate wave function is optimized using only these 4 VB structures only, the CCSD(T) equilibrium geometry is unstable by 5.4$\pm$0.9 kcal/mol. This shows that these four structures alone are not sufficient to describe the particular inter-fragment bonding in DTCNE, despite the sum of the weights of these 4 VB structures amounting to 72.6\% of the total weight of the full-basis 10-structure wave function. Structures 5 and 6, on their side, describe a situation with respectively two and four active electrons on each fragment. Therefore, adding structures 5 and 6 to the wave function enables electron fluctuation between the two monomers, and thus enables intra-fragment 2-center 3-electron bonding to take place between the two facing central carbon atoms of the two monomers. Adding these two extra charge-shift inter-fragment structures (which, altogether, sum to a 20.2\% weight in the full 10-structure VB-VMC wave function) stabilizes the dimer such that the energy of the dimer interaction is -5.2$\pm$0.9 kcal/mol, and therefore it is now stable in this geometry which is otherwise unfavorable as far as the electrostatic interaction is concerned. Therefore, structures 5 and 6 are shown to be crucial to explain the stability of the DTCNE pancake bonding system. Adding the remaining four minor charge-transfer VB structures (not shown in Figures~\ref{fig:TCNE_scheme2} and ~\ref{fig:TCNE_scheme3}) then slightly stabilizes the system further by an additional 1.7 kcal/mol. Improving the treatment of dynamical correlation by projecting the optimized VB-VMC wave function onto the FN ground state using the DMC algorithm gives an final interaction energy of -9.7$\pm$0.9 kcal/mol, which is in good agreement with the CCSD(T) and the MCQDPT/CASSCF values.

\begin{figure}[t]
    \centering
    \includegraphics[width=15cm]{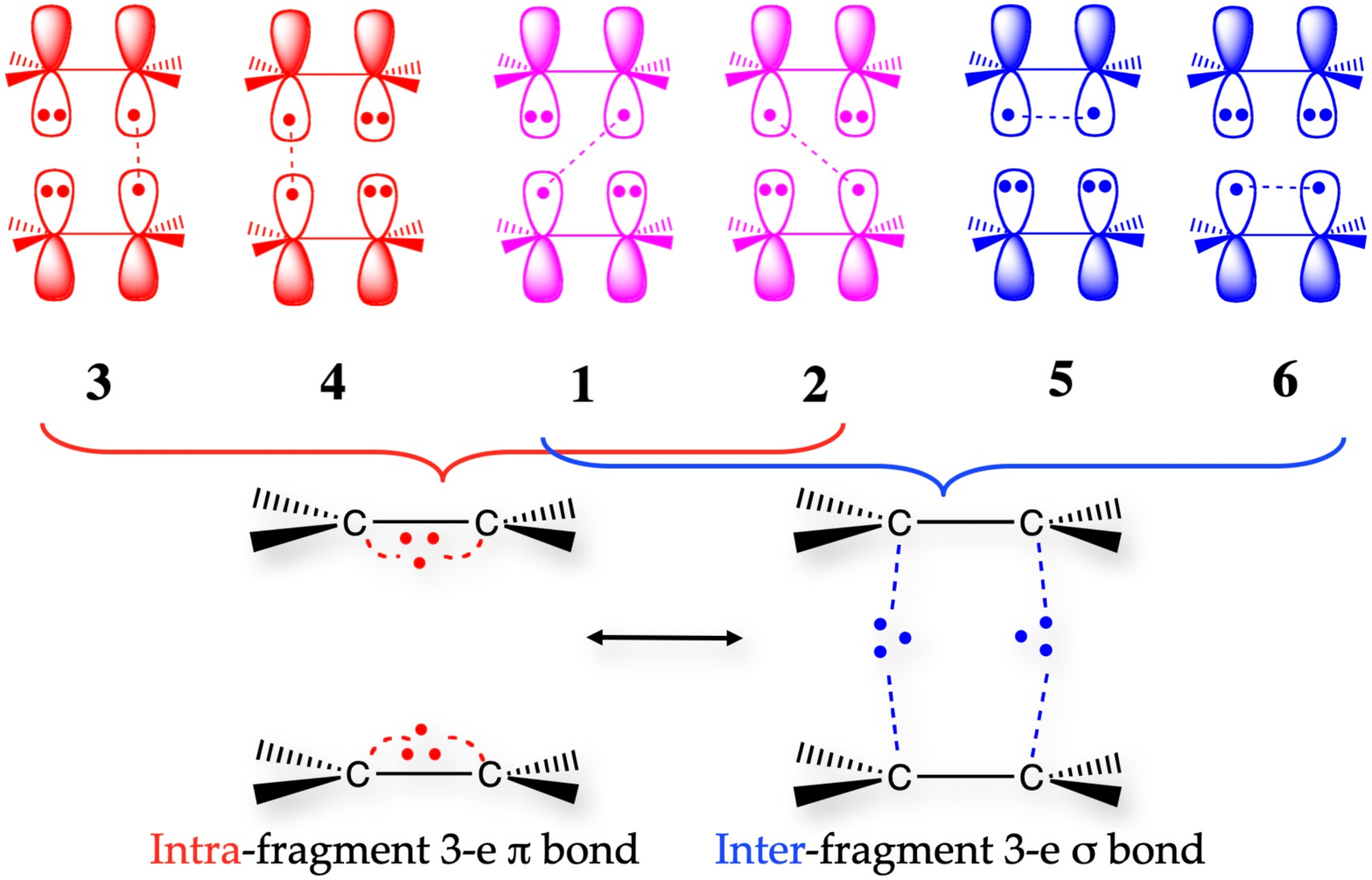}
    \caption{The VB model arising from the accurate VB-QMC calculations for the chemical component of bonding in DTCNE, as a combination of intra-fragment and inter-fragment 2-center 3-electron $\pi$ bonds.}
    \label{fig:TCNE_scheme4}
\end{figure}

Overall, the reading of the different computed VB-QMC calculations, and the key role of structures 5 and 6 in obtaining a stable DTCNE dimer in this geometry, could be summarized by Figure~\ref{fig:TCNE_scheme4}: the particular chemical bonding implying the six $\pi$-type electrons in the four central carbons corresponds to a combination of intra-fragment 2-center 3-electron $\pi$-bonds, described by the mixing of structures 1-4, and stabilizes only the separate monomers, with inter-fragment 2-center 3-electron $\pi$-bonds, enabled when structures 5 and 6 are added (arising from the resonance stabilization between structures 3-6) and correspond to the particular chemical interaction taking place between fragments in DTCNE. This component of bonding determines the inter-fragment distance and is responsible of the short carbon-carbon bonds, because the overlap between the optimized active orbitals facing each other between the two monomers is found to be 0.156, almost exactly the optimal overlap of 0.17 for 2-center 3-electron bonds. This ``charge-shift'' component of bonding adds up to the favorable electrostatic component in the equatorial configuration (left geometry in Figure~\ref{fig:TCNE_scheme1}) and to the dispersion interaction to lead to the overall large stabilization energy with respect to separate monomers in the minimum geometry. However, the 2-center 3-electron interaction is shown to be key to understand and account altogether for: the particular properties (in particular, the carbon-carbon inter-fragment distances) of these dimers, the large both static and dynamical correlation character, and the stability of the axial configuration (right geometry in Figure~\ref{fig:TCNE_scheme1}). 
Last, we shall mention that the importance of the cyano substituents was investigated by performing a similar VB-QMC study on the axial sodium ethylene dimer, (Na$^+$)$_2$(H$_4$C$_2$)$_2^{2-}$. At the counter-poise corrected CCSD(T) level, only a metastable configuration of the ethylene dimer was found with an intermolecular separation of 2.75{\AA},  with a barrier to dissociation of only 4.7 kcal/mol, and an overall exothermic dissociation by 10.9 kcal/mol. Due to the carbon atoms of ethylene being less electronegative than those of TCNE, the intramolecular overlap was stronger for the ethylene anion and the intermolecular overlap was reduced. This weakened the intermolecular 3-electron bonds hence the reduced stability of the ethylene dianionic dimer. This weakened intermolecular 3-electron bond character manifests itself in the reduced weights for the key VB structures 5 and 6, and in a similar exothermicity of 7.9$\pm$0.4 kcal/mol for the metastable configuration at the VB-VMC level using 6 VB structures. Hence, the electronegative substituents are essential for stabilizing stacked pancake anionic structures by stabilizing structures 5 and 6 that display two negative charges on the central atoms of the same fragment. By stabilizing structures 5 and 6, the cyano substituents then allow these structure to more strongly mix with the main VB structures 1-4, leading to a stronger 3-electron inter-fragment bonding that provides such molecular systems with enhanced stability and particular electronic properties.

\subsection{The XeF$_2$ prototype}

Ab initio VB theory coupled to QMC methods were successfully applied to gain detailed insight into the origin of the amazing stability of the XeF$_2$ molecule,\cite{XeF2-Hoppe-AngewChem-1962} which is a prototype of hypervalent compounds. In particular, Bra\"ida and Hiberty~\cite{XeF2-NaturChem-2013} showed that all models of hypervalency are insufficient to explain the great stability of XeF$_2$, and that charge-shift bonding in this molecule has the dominant role. Besides, in this study, the potential contribution of d orbitals to bond energy has been for the first time clearly ruled out based on a tailored VB treatment using the VB-VMC methodology. The essentials of this study is accounted for in the following.

Hypervalency in XeF$_2$ and its isoelectronic analogous (XeCl$_2$, KrF$_2$, ReF$_2$, ClF$_3$, SF$_4$,...) has been continuously in focus of theoretical chemists for almost a century~\cite{Pauling-1931-nature}. The very first model of hypervalency was proposed by Pauling~\cite{Pauling-1931-nature} by employing the sp$^3$d hybridization which assumes promotion of electrons into vacant d orbitals. This proposition was later reexamined, and based on population analysis it was argued that the high-lying d orbitals in hypervalent species have not large enough electronic occupations to be considered to play a leading role in bonding in these species.~\cite{d-role-Reed-JACS-1986} However, the approaches based on delocalized molecular orbitals (MO) did not allow to clearly separate the role of d functions as polarization functions or as bonding functions, so the question remained open.

\begin{figure}[t]
    \centering
    \includegraphics[width=8cm]{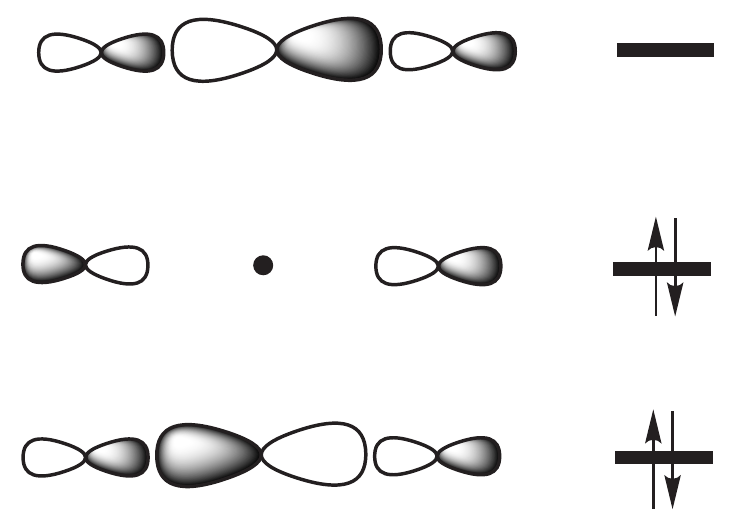}
    \caption{The Rundle–Pimentel orbital model for 3c–4e hypervalent bonding in XeF$_2$.}
    \label{fig:Rundle_Pimente_model}
\end{figure}

Pimentel and Rundle~\cite{Pimentel-JCP-1951, Rundle-JACS-1951} elaborated a simple MO-based model by introducing a three-centre–four-electron (3c–4e) scheme (Figure ~\ref{fig:Rundle_Pimente_model}). In the case of XeF$_2$, from three p$_z$ orbitals one can obtain the set of MOs of which only the first two are occupied, giving a total bond order of 0.5. The VB model of the hypervalency in XeF$_2$ and related species was proposed by Coulson~\cite{Coulson-XeF2-1964} by employing the resonance of a few VB structures:
\begin{equation}
\text{F-Xe}^{+}\text{F}^{-} \longleftrightarrow  \text{F}^{-}\text{Xe}^{+}\text{-F} \longleftrightarrow \text{F}^{-}\text{Xe}^{2+}\text{F}^{-}.
\end{equation}

It was shown~\cite{Coulson-XeF2-1964} that the Coulson model represents a VB-based way to describe the 3c–4e bonding as originally proposed by Pimentel and Rundle. Although, the Coulson–Rundle–Pimentel (CRP) model was accepted by a wide number of chemists, it fails to explain why F$_{3}^{-}$ is stable but H$_{3}^{-}$ is not. In addition, the CRP model does not provide any quantitative description of the hypervalent bonds.

\begin{figure}[t]
    \centering
    \includegraphics[width=15cm]{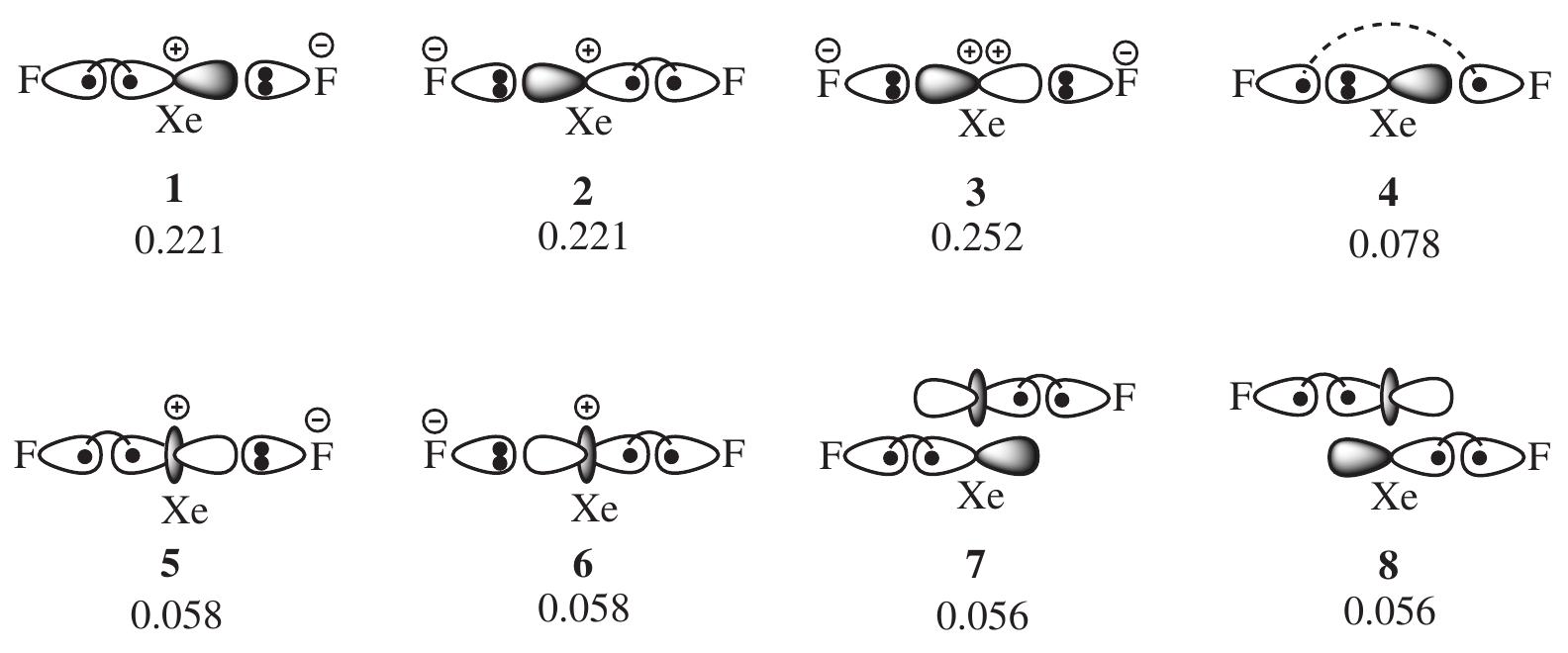}
    \caption{VB structures of XeF$_2$ and their Chirgwin-Coulson weights, calculated at the VB-VMC level.}
    \label{fig:XeF2_VB}
\end{figure}

Bra\"ida and Hiberty~\cite{XeF2-NaturChem-2013} performed a detailed VB-QMC study of the bonding in XeF$_2$, as a prototype of many hypervalent compounds. In these calculations a polarized triple-zeta basis set with relativistic pseudopotentials (indicated as ps-VTZ) was employed. The VB description of the active space of XeF$_2$ was based on the eight VB structures presented in Figure~\ref{fig:XeF2_VB}. In order to better understand the role of the 5d$_{z^2}$ orbitals of Xe the employed VB structures were gathered into two groups. In VB structures 1-4 the active electrons of Xe are in the axial p$_z$ orbital, whereas in structures 5-8 at least one covalent bond involves the 5d$_{z^2}$ orbital of Xe. Structures 7 and 8 characterize the direct participation of 5d$_{z^2}$ to the F-Xe bond, following the sp$^{3}$d hybridization scheme of Pauling. Accordingly, the contribution of structures 7 and 8, in terms of weights or in terms of stabilization energy, is a direct measure of the importance of the Pauling expanded octet model. 

The calculated relative energies of XeF$_2$ with respect to its separate atoms are presented in Table~\ref{tab:Atom_Ene_XeF2}. At the VB-VMC level, the calculated atomization energy is 48.7 kcal/mol, better than the CCSD(T) value in the same basis set, but somewhat smaller than the experimental value and the complete-basis-set (CBS) extrapolated CCSD(T) value, reflecting the necessity of very large basis sets to obtain accurate energetics for this molecule. It shall be mentioned that the SD-BOVB level has led to a significantly smaller atomization energy than both the CCSD(T) and VB-VMC levels in this very demanding test system. The VB-DMC level corrects for most of the remaining basis-set deficiencies and correlation effects missing in the trial wave function, and provides an atomization energy in excellent agreement with the experimental value.

\begin{table}
\caption{Atomization energy of XeF$_2$, as calculated by
CCSD(T), VB-VMC, and VB-DMC methods using the ps-VTZ basis set.}
\label{tab:Atom_Ene_XeF2}
\begin{tabular}{|c|c|} 
 \hline
       & Relative energy (kcal/mol) \\
 \hline
 CCSD(T)   & 40.1 \\
 VB-VMC$^a$    & 48.7 \\
 VB-DMC$^a$    & 60.6 \\
 CCSD(T)/CBS$^b$      & 63.7 \\
 Exp.       & 62.2$^c$, 63.4$^d$ \\
 \hline
 \multicolumn{2}{l}{$^a$From Ref.~\onlinecite{XeF2-NaturChem-2013}.}\\[-0.2cm]
 \multicolumn{2}{l}{$^b$CCSDT(T) extrapolated to the complete-basis-set (CBS) limit.}\\[-0.2cm]
 \multicolumn{2}{l}{$^c$Experimental value from Ref.~\onlinecite{XeF2-BDE-weinstock-1966}.}\\[-0.2cm]
 \multicolumn{2}{l}{$^d$Experimental value from Ref.~\onlinecite{XeF2-BDE-berkowitz-1971}.}
\end{tabular}
\end{table}
As can be seen from Figure~\ref{fig:XeF2_VB} the three VB structures of the CRP model are largely dominant. These are covalent structures 1 + 5 and 2 + 6, and ionic structure 3, which together represent 81\% of the electronic state. Finally, structures 7 and 8 together contribute only 11.2\% to the wave function, thereby demonstrating the marginal contribution of sp$^{3}$d hybridization. The stabilization energy contributed by structures 7 and 8 was estimated by comparing the atomization energies calculated with and without inclusion of structures 7 and 8 in the VB set. Table~\ref{tab:diabatic_Ene} displays the energies of some individual VB structures (or combinations of them) for XeF$_2$ relative to its separate atoms. It was found that removing structures 7 and 8 from the wave function results in an energy increase of only 7.2 kcal/mol, a value that can be considered a quantitative measure of the contribution of sp$^{3}$d hybridization to the stability of XeF$_2$. These results demonstrate, in a more quantitative way, the definite superiority of the CRP model over the expanded octet proposal.

\vspace{5mm}

\begin{table}
\caption{Energies of some diabatic states of XeF$_2$, as
calculated by the VB-VMC method (from Ref.~\onlinecite{XeF2-NaturChem-2013}).}
\label{tab:diabatic_Ene}
\begin{tabular}{ |c|c| } 
 \hline
 Set of VB structures & Relative energy (kcal/mol) \\
 \hline
 1+5 or 2+6                                 & 104.3 \\
 3                                           & 78.8  \\
 1+2+5+6                                     & 21.4  \\
 1+2+4+5+6                                   & 2.8   \\
 1+2+3+4+5+6                                 & -41.5 \\
 1+2+3+4+5+6+7+8                              & -48.7 \\
 \hline
\end{tabular}
\end{table}
The energies of the diabatic states given in Table~\ref{tab:diabatic_Ene} also provide the extent to which the resonance energies participate to the stability of XeF$_2$. It is manifest that both covalent (1+5 or 2+6) and ionic (3) structures of the CRP model are unbound. On the other hand, mixing the two covalent structures (1+2+5+6) results in a stabilization of 82.9 kcal/mol. This large resonance energy in XeF$_2$ arises from the charge-shift character of the corresponding covalent bonding. Another expression of charge-shift bonding in XeF$_2$ can be seen in the large resonance energy (44.3 kcal/mol) arising from the mixing of the ionic structure with the combination of covalent structures. This covalent-ionic resonance energy is close to the total bonding energy at this level of calculation, which is the very definition of a charge-shift-bonded system. Thus, charge-shift bonding is a key feature of hypervalency in XeF$_2$, and without the large resonance energies that are attached to this type of bonding the molecule would be unstable. 

The performed high-level VB-QMC calculations provided the complete explanation of the strong stability of XeF$_2$ as a combination of the CRP model with the occurrence of charge-shift bonding. In addition, this study showed the importance of the diionic structure 3 which is often omitted in papers or textbooks. 

The overall conclusion of this analysis is that the CRP VB-like four-structure description of the electron-rich XeF$_2$ hypervalent system is an effective and faithful model, however it is the charge-shift mechanism which is the glue that makes this system stable, through the large resonance stabilization arising from the mixing of the four dominant VB structures. Based on this detailed and accurate ab initio VB analysis, a ``rule of thumb'' has been proposed in a continuation study~\cite{BraRibHib-CEJ-14} to identify the potentially stable electron-rich hypervalent systems. Namely, the stable systems will be those where (1) the peripheric atoms are electron-rich and electronegative atoms bear lone pairs (fluorine atoms being the best candidate), a condition to obtain large resonance energy stabilization from an effective charge-shift mechanism, and (2) where the central atom has small first and second ionization potentials, a condition leading to a diionic structure low enough in energy to mix strongly with the other three, thus leading to a large enough charge-shift resonance energy to obtain a stable hypervalent compound.

As a final conclusion, this study also illustrated that a classical VB method that involves a sophisticated treatment of electron correlation such as the VB-VMC method, can prove to be a reliable tool to reach new and ground solid chemical insight, because the qualitative models are inferred from the direct analysis of a highly correlated VB wave function which encompasses the essential physics of the system, as proved by the accurate properties reproduced such as atomization energies, and therefore such derived models are devoid of the artifact of the usual semi-quantitative analysis based on more approximate wave functions or on gross simplifications of high-level wave functions.

\section{Conclusion}

    As we have demonstrated in this chapter, VB-VMC and VB-DMC methods are correlated classical VB methods which are capable of achieving high accuracy in calculations of energies and energy differences of molecular systems with complex electronic structures. Moreover, the VB-VMC method is capable of retaining the full chemical interpretability of the traditional ab initio classical VB methods, with, in particular, the ability of calculating reliable weights of VB structures, and optimizing separate VB wave functions corresponding to individual diabatic states which provide access to accurate resonance energies. The merits of supplementing VB-VMC with VB-DMC, when highly accurate energy differences are required, have been illustrated in several applications in this chapter. As compared with MO-based wave functions, it has been shown that in some cases, such as the N $\to$ V excitation in ethylene or DTCNE pancake bonding systems, a correlated classical VB calculation such as VB-VMC can lead to an extremely compact correlated wave function capable of nearly reaching chemical accuracy. As far as QMC calculations are concerned, the compactness of such VB-based trial wave functions allow for faster QMC calculations, since the locality of active classical VB orbitals also contributes to speeding up the calculations as compared with Jastrow-Slater multideterminant expansions based on delocalized orbitals. As far as interpretative-oriented ab initio classical VB studies are concerned, because the QMC algorithms remain the same whether the orbitals are orthogonal or not, optimization of rather large non-orthogonal determinantal expansions at the VB-VMC level is possible, providing an alternative approach to the largely used BOVB method which is limited to a handful of structures because of the unfavorable scaling of the standard algorithm used in BOVB. Of course, the limitation of this VB-QMC approach, which is the same as in all QMC methods, is their massive computational requirements, which however is less of an issue on massively parallel computing devices, because of the inherently parallel nature of QMC algorithms. As a general conclusion, we therefore believe that the VB-QMC approach constitutes a useful addition to the family of VB methods, and more generally to the toolbox of theoretical chemists, and shall gain in popularity in the future.


\end{document}